\documentclass[usenatbib]{mn2e}
\usepackage{aasmacros, amsmath, amssymb, color, graphicx, multirow, mymacro}
\defcitealias{TinWetCon11}{Paper~I}
\defcitealias{WetTinCon12a}{Paper~II}
\defcitealias{WetTinCon12b}{Paper~III}
\defcitealias{WetTinCon13a}{Paper~IV}
\defcitealias{WetTinCon13b}{Paper~V}
\defcitealias{Tin12}{Tinker et al., in prep.}
\defcitealias{Wet12}{Wetzel et al., in prep.}

\begin{document}
\title[Satellite galaxy star formation histories]{Galaxy evolution in groups and clusters: satellite star formation histories and quenching timescales in a hierarchical Universe}
\author[Wetzel, Tinker, Conroy \& van den Bosch]{Andrew R. Wetzel${}^1$, Jeremy L. Tinker${}^2$, Charlie Conroy${}^3$, and Frank C. van den Bosch${}^1$\\
$^{1}$Department of Astronomy, Yale University, New Haven, CT 06520, USA\\
$^{2}$Center for Cosmology and Particle Physics, Department of Physics, New York University, New York, NY 10013, USA\\
$^{3}$Harvard-Smithsonian Center for Astrophysics, Cambridge, MA 02138, USA
}
\date{June 2012}
\pagerange{\pageref{firstpage}--\pageref{lastpage}} \pubyear{2012}
\maketitle
\label{firstpage}

\begin{abstract}
Satellite galaxies in groups and clusters are more likely to have low star formation rates (SFR) and lie on the `red-sequence' than central (`field') galaxies.
Using galaxy group/cluster catalogs from the Sloan Digital Sky Survey Data Release 7, together with a high-resolution, cosmological \tit{N}-body simulation to track satellite orbits, we examine the star formation histories and quenching timescales of satellites of $\mstar > 5 \times 10 ^ {9} \msun$ at $z \approx 0$.
We first explore satellite infall histories: group preprocessing and ejected orbits are critical aspects of satellite evolution, and properly accounting for these, satellite infall typically occurred at $z \sim 0.5$, or $\sim 5 \gyr$ ago.
To obtain accurate initial conditions for the SFRs of satellites at their time of first infall, we construct an empirical parametrization for the evolution of central galaxy SFRs and quiescent fractions.
With this, we constrain the importance and efficiency of satellite quenching as a function of satellite and host halo mass, finding that satellite quenching is the dominant process for building up all quiescent galaxies at $\mstar < 10 ^ {10} \msun$.
We then constrain satellite star formation histories, finding a `delayed-then-rapid' quenching scenario: satellite SFRs evolve unaffected for $2 - 4 \gyr$ after infall, after which star formation quenches rapidly, with an e-folding time of $< 0.8 \gyr$.
These quenching timescales are shorter for more massive satellites but do not depend on host halo mass: the observed increase in satellite quiescent fraction with halo mass arises simply because of satellites quenching in a lower mass group prior to infall (group preprocessing), which is responsible for up to half of quenched satellites in massive clusters.
Because of the long time delay before quenching starts, satellites experience significant stellar mass growth after infall, nearly identical to central galaxies.
This fact provides key physical insight into the subhalo abundance matching method.
\end{abstract}


\section{Introduction}

Observations have long shown that galaxies in denser regions are more likely to have low star formation rates (SFR), lie on the red sequence, and exhibit more evolved (elliptical) morphologies than similar mass galaxies in less dense regions, from massive galaxies in clusters \citep{Oem74, DavGel76, Dre80, DreGun83, PosGel84, BalMorYee97, PogSmaDre99} to the lowest mass satellites in the Local Group \citep{Mat98}.

Large-scale galaxy surveys, such as the Sloan Digital Sky Survey \citep [SDSS;][]{SDSS}, have enabled detailed examinations of the correlations between these galaxy properties and their environment at $z \approx 0$ \citep[see][for a recent review]{BlaMou09}.
Several such early works showed that galaxy SFR/color depends on small-scale ($\lesssim 1 \mpc$) environment, with little-to-no additional dependence on larger scale environment \citep{HogBlaBri04, KauWhiHec04, BlaEisHog05}.
More physically, this environmental dependence has been shown to result from satellite galaxies and the properties of their host dark matter halo \citep{WeivdBYan06a, BlaBer07, WilZibBud10, TinWetCon11}, where `satellite' galaxies are all those that are not the massive `central' galaxy at the core of the host halo.

These results are physically meaningful, given that the virial radius corresponds to a physical transition from the low-density `field' environment to a high-density, virialized region.
After a satellite falls into a host halo, the strong gravitational tidal forces prevent the satellite's (sub)halo from accreting dark matter and also strip mass from the subhalo from the outside-in \citep[e.g.,][]{DekDevHet03, DieKuhMad07}.
Additionally, if the host halo is massive enough to host a stable, virial accretion shock \citep{DekBir06}, then its thermalized gas also can heat and strip any extended gas in the subhalo \citep{BalNavMor00, KawMul07, McCFreFon08}.
Therefore, satellites (eventually) experience reduced gas cooling/accretion rates onto their disc after infall, a phenomenon known as `strangulation' or `starvation' \citep{LarTinCal80}.
More drastically, in the extreme case of both high gas density and satellite velocity, ram-pressure can strip cold gas directly from the disc \citep{GunGot72, AbaMooBow99, ChuvGoKen09}.
The dense collection of galaxies in a host halo also allows for the possibility of strong gravitational interactions with neighboring galaxies, known as `harassment' \citep{FarSha81, MooLakKat98}, and satellites can merge with one another \citep{MakHut97, AngLacBau09, WetCohWhi09a, WetCohWhi09b, WhiCohSmi10, Coh12}.
All of these mechanisms are expected to play some role in quenching satellite star formation, though their importance, particularly as a function of host halo mass, remains in debate.

To constrain satellite quenching processes, many works have examined the SFRs/colors of satellites in groups/clusters in detail at $z \approx 0$ \citep[e.g.,][]{BalNavMor00, EllLinYee01, DePColPea04, WeivdBYan06a, BlaBer07, KimSomYi09, vdBAquYan08, HanSheWec09, KimSomYi09, PasvdBMo09, vdLWilKau10, PreBalJam11, PenLilRen11, WetTinCon12a, WooDekFab13}.
In general, these works found that the fraction of satellite that are quiescent/red, $\fsatq$, depends primarily, and independently, on three quantities: $\fsatq$ (1) increases with satellite mass, (2) increases with the mass of the host halo, and (3) increases toward halo center.
Trend (1) is caused, at least partially, by the underlying dependence on stellar mass set by central galaxies prior to infall.
Trend (2) is sometimes interpreted as satellites being quenched more rapidly in more massive host halos, but the hierarchical nature of halo growth, namely, the possibility of quenching as a satellite in a lower mass halo prior falling into a more massive halo (`group preprocessing') complicates this interpretation \citep[e.g.,][]{ZabMul98, McGBalBow09}.
Finally, trend (3) implies an evolutionary trend, because a satellite's halo-centric radius negatively correlates with its time since infall \citep[e.g.,][]{GaoWhiJen04, DeLWeiPog12}.
Similar satellite trends persists out to at least $z \sim 1$ \citep[e.g.,][]{CucIvoMar06, CooNewCoi07, GerNewFab07, TraSaiMou09, PenLilKov10, McGBalWil11, GeoLeaBun11, MuzWilYee12}.

Several works have gone beyond simple satellite SFR/color cuts to examine observationally the nature of the full SFR/color distribution.
These works have shown that the color \citep{BalBalNic04, Ski09} and SFR \citep{BalEkeMil04, McGBalWil11, PenLilRen11, WetTinCon12a, WooDekFab13, WijHopBro12} distribution of galaxies is strongly bimodal across all environmental/halo regimes, and the SFR/color of actively star-forming/blue galaxies does not vary with any environmental measure.
As noted in many of the above works, these results imply that the environmental process(es) takes considerable time (several Gyrs) to affect satellite SFR.

In this paper, we seek to use the aforementioned observational trends, which we presented in detail in \citet{WetTinCon12a}, to quantify---in a robust, statistical manner---the star formation histories and quenching timescales of satellite galaxies at $z = 0$ across a wide range of both satellite and host halo masses.
Understanding satellite quenching mechanisms and the timescales over which they operate is important for elucidating the physical processes that occur in groups and clusters, but also for a comprehensive understanding of galaxy evolution overall, given that satellites constitute $\sim 1 / 3$ of all low-mass galaxies \citep[e.g.,][]{YanMovdB07}.
Satellite galaxies also provide unique laboratories for examining gas depletion and its relation to star formation because, unlike central galaxies, satellites are thought not to accrete gas from the field after infall.
Furthermore, because satellites are significantly more likely to lie on the red sequence, many methods for identifying galaxy groups/clusters rely on selecting red-sequence galaxies \citep[e.g.,][]{GlaYee00, KoeMcKAnn07a}, so a detailed understanding of the systematics of these methods requires characterizing the timescale over which satellites migrate onto the red sequence after infall and how this timescale depend on host halo mass and redshift.

Many works have investigated satellite SFR evolution and quenching through the use of semi-analytic models (SAMs) applied to cosmological \tit{N}-body simulations.
In one early example, \citet{BalNavMor00} modeled satellite SFR as declining exponentially after infall on a cold gas consumption timescale of a few Gyrs to account for radial gradients of average SFR and color in clusters.
Many SAMs assumed that a satellite subhalo's hot gas is stripped instantaneously as it passes within the host halo's virial shock, but this scenario quenches star formation too rapidly; only models that remove/strip gas more gradually produce realistic quiescent fractions \citep{WeivdBYan06b, FonBowMcC08, KanvdB08, BooBen10}, but, in general they have difficulty in correctly reproducing the full SFR distribution.
Though, \citet{WeiKauvdL10} recently implemented a modification of the SAM of \citet{DeLBla07} in which the diffuse gas around a satellite galaxy is stripped at the same rate as its host dark matter subhalo (10 - 20\% loss per Gyr), showing that this modification produces a satellite SSFR distribution that broadly is in agreement with observations.
However, understanding the results of SAMs is complicated by the fact that they do not completely accurately model the evolution of central galaxies, so they do not provide fully accurate initial conditions at infall for satellites.
Relatedly, a few works have examined satellite SFR evolution in cosmological hydrodynamic simulations of galaxy groups \citep[e.g.,][]{FelCarMay11}, arguing that quenching occurs largely through the lack of gas accretion and (to a lesser degree) gas stripping after infall.

Instead of attempting fully to forward-model all of the relevant physical processes for satellites, our approach is to parametrize satellite star formation histories and constrain their quenching timescales in as much of an empirical manner as possible.
We start with detailed measurements of the SFRs of satellites at $z = 0$ from our SDSS group catalog that we presented in \citet{WetTinCon12a} and review in \S\ref{sec:method}.
We also describe our cosmological $N$-body simulation, which we use to create a mock group catalog to compare our models to observations robustly.
With this simulation, we explore the infall times of satellites in \S\ref{sec:infall-time}.
We then develop an accurate, empirical parametrization for the initial SFRs of satellites at their time of infall in \S\ref{sec:sfr_at_infall}.
Having accurate initial SFRs of satellites and measurements of their final SFRs at $z = 0$, we examine the importance and efficiency of satellite quenching in \S\ref{sec:quench_importance_efficiency} and their star formation histories and quenching timescales in \S\ref{sec:sfr-evol_sat}.
With this, we then examine where satellites were when they quenched in \S\ref{sec:where_quench} and their stellar mass growth in \S\ref{sec:m-star_growth}.
Finally, we discuss the implications of our results for subhalo abundance matching in Appendix \ref{sec:mass_growth_sham}.

This paper represents the third in a series of four.
In \citet{TinWetCon11}, hereafter \citetalias{TinWetCon11}, we described our SDSS galaxy sample, presented our method for identifying galaxy groups/clusters, and showed that central and satellite galaxy quiescent fractions are essentially independent of the large-scale environment beyond their host halo.
In \citet{WetTinCon12a}, hereafter \citetalias{WetTinCon12a}, we used our SDSS group catalog to examine in detail the SFR distribution of satellite galaxies and its dependence on stellar mass, host halo mass, and halo-centric radius, finding that the SFR distribution is strongly bimodal in all regimes.
Based on this, we argued that satellite star formation must evolve in the same manner as central galaxies for several Gyrs after infall, but that once satellite quenching starts, it occurs rapidly.
In \citet{WetTinCon13a}, hereafter \citetalias{WetTinCon13a}, we will examine quenching in galaxies \tit{near} groups and clusters, focusing on `ejected' satellites that passed within a more massive host halo but have orbited beyond the virial radius.
We will show that these ejected satellites can explain essentially all trends for star formation quenching in galaxies beyond the the virial radius of groups/clusters.
Finally, in \citet{WetTinCon13b}, hereafter \citetalias{WetTinCon13b}, we will use the detailed orbital histories from our simulation to constrain the physical mechanisms responsible for satellite quenching.

For clarity, we outline some nomenclature.
We refer to galaxies as `quiescent' in an observational sense: having low SFR but without regard to how or when SFR faded.
By contrast, we refer to satellite `quenching' in our models as the physical process of SFR fading rapidly below the quiescence threshold, under the ansatz that once a satellite is quenched it remains so indefinitely.
Our galaxy group catalog refers to `group' in a general sense, as a set of galaxies that occupy a single host halo, regardless of its mass, and we will use `(host) halo' as a more general term for group or cluster.
Finally, we cite all masses using $h = 0.7$ for the Hubble parameter.

\section{Methods} \label{sec:method}

In this section, we first briefly describe our galaxy sample and group-finding algorithm.
(For full details, see \citetalias{TinWetCon11} for our galaxy sample and group-finding algorithm, and \citetalias{WetTinCon12a} for our SFR measurements.)
We then describe our simulation, subhalo finding and tracking, and methodology for making galaxy and group catalogs in the simulation.

\subsection{SDSS Galaxy Catalog}

Our galaxy sample is based on the NYU Value-Added Galaxy Catalog \citep{BlaSchStr05} from SDSS Data Release 7 \citep{AbaAdeAgu09}.
Galaxy stellar masses are from the {\tt kcorrect} code of \citet{BlaRow07}, assuming a \citet{Cha03} initial mass function (IMF).
We construct two volume-limited samples of all galaxies with $M_r - 5 \log(h) < -18$ and $-19$\footnote{
In \citetalias{WetTinCon12a}, we wrote this as $M_r < -18$ and -19, but we did not assume a value for $h$ in calculating magnitudes.}
, which go out to $z = 0.04$ and $0.06$, from which we identify stellar mass completeness limits of $5 \times 10 ^ {9}$ and $1.3 \times 10 ^ {10} \msun$, respectively.
Combining these samples leads to an overall median redshift of $z = 0.045$, though we will indicate this as $z = 0$ for brevity.

For our galaxy star formation metric we use specific star formation rate, $\ssfr = \sfr / \mstar$, based on the current release\footnote{
\tt http://www.mpa-garching.mpg.de/SDSS/DR7/} 
of the spectral reductions of \citet{BriChaWhi04}, with updated prescriptions for active galactic nuclei (AGN) contamination and fiber aperture bias corrections following \citet{SalRicCha07}.
These SSFRs are derived primarily from emission lines (mostly $\halpha$), but in cases of strong AGN contamination or no measurable emission lines, the SSFRs are inferred from $\dnfk$.
Roughly, $\ssfr \gtrsim 10 ^ {-11} \yrinv$ are based almost entirely on $\halpha$, $10 ^ {-12} \lesssim \ssfr \lesssim 10 ^ {-11} \yrinv$ are based on a combination of emission lines, and $\ssfr \lesssim 10 ^ {-12} \yrinv$ are based almost entirely on $\dnfk$ and should be considered upper limits to the true value \citep{SalRicCha07}.
The use of spectroscopically derived SSFRs is critical for our analysis because dust reddening causes simple red/blue color cuts to overestimate the quiescent fraction by up to 50\%, particularly at lower mass (see Fig.~1 in \citetalias{TinWetCon11}).

\subsection{SDSS Group Catalog} \label{sec:group_catalog_sdss}

Motivated by the paradigm that all galaxies reside in host dark matter halos, we identify groups of galaxies that occupy the same host halo and their halo properties through a modified implementation of the group-finding algorithm of \citet{YanMovdB05a, YanMovdB07}.
For our group catalog, we define dark matter host halos such that the mean matter density interior to the virial radius is 200 times the mean background matter density: $\mthm = 200 \bar{\rho}_{\rm m} \frac{4}{3} \pi \rthm^3$.
We place galaxies into groups through an iterative procedure outlined in \citetalias{TinWetCon11}, using the $\mstar > 10 ^ {9.7} \msun$ sample at $z < 0.04$ and the $\mstar > 10 ^ {10.1} \msun$ sample at $0.04 < z < 0.06$.
We assign dark matter halo masses to groups by matching the abundance of halos above a given dark matter mass to the abundance of groups above a given total stellar mass: $n(> M_{\rm vir, halo}) = n(> M_{\rm star, group})$.
Here, we use the host halo mass function from \citet{TinKraKly08}, based on a flat, $\Lambda$CDM cosmology of $\Omega_{\rm m} = 0.27$, $\Omega_{\rm b} = 0.045$, $h = 0.7$, $n_{\rm s} = 0.95$ and $\sigma_8 = 0.82$, consistent with a wide array of observations \citep[e.g.,][and references therein]{KomSmiDun11}.
Every group contains one `central' galaxy, which by definition is the most massive, and can contain any number (including zero) of less massive `satellite' galaxies.

We define a group's center by the location of its most massive galaxy.
However, in reality the most massive galaxy is not always the one closest to the minimum of a halo's potential well, particularly at high halo mass \citep{SkivdBYan11}.
This effect arises largely because of the shallowness of the $\mstar - \msubhalo$ relation at high mass, such that any non-negligible scatter in this relation leads to a significant probability that if a less massive halo falls into a more massive halo, the central galaxy of the less massive halo has higher stellar mass than that of the more massive halo.
However, these are typically cases in which two galaxies in a halo have similarly high mass, and in this regime almost all galaxies exhibit quiescent SSFRs regardless of central versus satellite demarcation.
Also, because we will assume realistic scatter (0.15 dex) in the $\mstar - \msubhalo$ relation in making our simulation group catalog (see \S\ref{sec:galaxy_catalog_sim}), this effect propagates into our model results as well.

Because of projection effects and redshift-space distortions, our groups inevitably contain interloping galaxies: some central galaxies are mis-assigned as satellites in higher mass halos (reducing group purity), and conversely some satellites are mis-assigned as central galaxies of lower mass halos (reducing group completeness).
As detailed in \citetalias{TinWetCon11}, an average of $\sim 10\%$ of galaxies are mis-assigned in this way \citep[see also][]{YanMovdB07}.
In this work, we apply the same group-finding algorithm to our simulation, as described below, allowing us to examine our theoretical results in `observational' space and correct for these effects.

\subsection{Simulation and Subhalo Tracking} \label{sec:simulation}

Our goal is to understand the SFR evolution of galaxies in groups/clusters and how this evolution connects with satellite infall times and lifetimes, which are governed by complex dynamical processes.
To this end, we require a cosmological simulation that both provides significant statistics across a broad range of host halo masses and can robustly track satellite evolution from first infall to final merging/disruption.
We employ a dissipationless, $N$-body simulation using the TreePM code of \citet{Whi02} with flat, $\Lambda$CDM cosmology of $\Omega_{\rm m} = 0.274$, $\Omega_{\rm b} = 0.0457$, $h = 0.7$, $n = 0.95$ and $\sigma_8 = 0.8$, nearly identical to the cosmology used in making our group catalog.
To achieve both high resolution and significant volume, the simulation evolves $2048 ^ 3$ particles in a $250 \hmpc$ box, with a particle mass of $1.98 \times 10^8 \msun$ and a Plummer equivalent smoothing of $2.5 \hkpc$.
Initial conditions are generated at $z = 150$ using second-order Lagrangian Perturbation Theory, with a displacement RMS of 38\% of the mean inter-particle spacing.
45 outputs are stored from $z = 10$ to 0, spaced evenly in $\ln(a)$, with output time spacings of 400 and $650 \myr$ at $z = 1$ and 0, respectively.
This same simulation was used in \citet{WhiCohSmi10}.

We identify `host halos' using the Friends-of-Friends (FoF) algorithm \citep{DavEfsFre85} with a linking length of $b = 0.168$ times the mean inter-particle spacing, which groups particles bounded by an isodensity contour of $\sim 100 \times$ the mean matter density.
($b = 0.2$ is often used, but it is more susceptible to joining together distinct, unbound structures.)
Note that this halo definition is different from the spherical overdensity definition used in making the SDSS group catalog, but we address this issue in \S\ref{sec:group_catalog_sim}.

Within host halos, we identify `subhalos' as overdensities in phase space through a 6-dimensional FoF algorithm (FoF6D), also described in \citet{WhiCohSmi10}.
Based on extensive experimentation, we use a configuration space linking length of 0.078 of the simulation's mean interparticle spacing and a velocity linking length of 0.368 of each host halo's 1D velocity dispersion.\footnote{
In rare cases, using these parameters leads to zero subhalos in a low-mass halo.
To avoid having halos with no subhalos, mostly for tracking purposes, we slowly increase the linking lengths in those halos, and we stop if we identify at least two subhalos.
This procedure does not affect halo mass ranges used in this work.}
Our tests show that our FoF6D implementation leads to good agreement with \textsl{SUBFIND} \citep{SprWhiTor01} for massive, well-resolved subhalos, but FoF6D is significantly more robust in tracking low-mass subhalos and those that pass close to halo center.
For both host halos and subhalos, we keep all objects with at least 50 particles, and we define its center and velocity by the position and velocity of its most bound particle.

We track host halos and subhalos across simulation outputs and build merger trees as described in \citet{WetCohWhi09b} and \citet{WetWhi10}, with slight modifications as given below.
We assign to each (sub)halo a unique `child' (sub)halo at the next simulation output, based on its 20 most bound particles.
We track subhalo histories across four consecutive outputs at a time because a subhalo can briefly disappear while passing through the center of its host halo or another subhalo.
In these cases, we ensure that the subhalo is identified at each output by creating `virtual' subhalos via interpolating the properties of the temporarily disappearing subhalos between outputs.
If a (sub)halo has multiple `parent' (sub)halos at the previous output, we identify the main parent as the most massive one (using $\mmax$ for subhalos, see below), and we use this main parent in tracking back a (sub)halo's history and identifying its main progenitor at an earlier time. 

We define a `central' subhalo as being the most massive subhalo in a newly-formed (has no parent) host halo at a given simulation output.
The central subhalo almost always corresponds to the object at the minimum of a halo's potential well.
A subhalo retains its `central' definition until falling into (specifically, becoming linked via FoF to) a more massive host halo, becoming a `satellite' subhalo.
Every sufficiently bound halo hosts one central subhalo at its core and can host zero, one, or multiple satellite subhalos, so these `central' and `satellite' definitions for subhalos closely reflect those of galaxies in the group catalog.

We assign to each subhalo a maximum mass, $\mmax$, motivated by the strong correlation of this quantity with galaxy stellar mass (see \S\ref{sec:group_catalog_sim}).
A subhalo's $\mmax$ is based on the maximum host halo mass that it ever had as a central subhalo (so it corresponds to FoF halo mass and not FoF6D subhalo mass).
For a central subhalo, $\mmax$ almost always corresponds to its current halo mass, the primary exception being those that have passed through a more massive halo and been ejected (see \S\ref{sec:infall_eject}).
For a satellite subhalo, $\mmax$ almost always corresponds to its halo mass sometime prior to infall, though not necessarily \tit{immediately} prior to infall because a satellite typically undergoes some mass stripping just prior to infall, arising from the strong tidal forces near a massive halo.
(The two effects above motivate our use of $\mmax$ instead of mass at infall, which was used in \citet{WetCohWhi09b, WetCohWhi09a} and \citet{WetWhi10}.)
Thus, a satellite's $\mmax$ remains fixed after infall, unless it merges with another satellite, in which case the resultant satellite is given the sum of its parents' $\mmax$ values.

As demonstrated in \citet{WetWhi10}, simulations at our high resolution scale can resolve and track massive satellite subhalos past the point at which the galactic stellar component that they host would (start to) be stripped, merge with the central galaxy, or otherwise be disrupted.
Not accounting for this galactic merging/disruption in \tit{N}-body simulations can lead to stronger small-scale clustering than seen in observations.
This effect is even more significant for our phase-space FoF6D subhalo finder, which tracks subhalos more robustly down to small halo radii.
To account for this effect, we use the subhalo merging/disruption scheme in \citet{WetWhi10} and remove subhalos with $M_{\rm bound} / \mmax < 0.007$, which provides good agreement with mass/luminosity-dependent clustering, satellite fractions, and luminosity functions in clusters for this simulation and FoF6D subhalo finder.
Thus, we properly resolve the orbital histories and infall times of all satellites in our sample ($\mstar > 5 \times 10 ^ {9} \msun$ corresponds to subhalo $\mmax > 3 \times 10 ^ {11} \msun$, see below).

\subsection{Simulation galaxy catalog} \label{sec:galaxy_catalog_sim}

Under the assumption that a galaxy resides at the center of each surviving dark matter subhalo, we assign stellar mass using subhalo abundance matching \citep[SHAM;][]{ValOst06, ConWecKra06}.
This method assumes a one-to-one mapping that preserves rank ordering between subhalo $\mmax$ (or maximum circular velocity) and galaxy $\mstar$, such that $n(> \mmax) = n(>\mstar)$, allowing one to assign $\mstar$ to subhalos empirically using an observed stellar mass function (SMF) that is recovered, by design.
SHAM has succeeded in reproducing many observed galaxy statistics, including spatial clustering, satellite fractions, cluster luminosity functions, and luminosity-velocity relations \citep{ConWecKra06, ValOst06, BerBulBar06, WanLiKau06, YanMovdB09, WetWhi10, TruKlyPri11}.
We note that, despite these successes, SHAM in its simplest incarnation, using $\mmax$ for both satellite and central subhalos, may not be fully accurate in assigning stellar mass to both satellite and central subhalos simultaneously, because there is some freedom in allowing satellites to follow a different relation \citep{NeiLiKho11, YanMovdB12, MosNaaWhi13}.
We discuss this issue further in Appendix \ref{sec:mass_growth_sham}.

For this work, we use the SMF from \citet{LiWhi09}, based on the same SDSS NYU-VAGC sample as our galaxy catalog, including the same $K$-correction and IMF.
We apply SHAM at a simulation output of $z = 0.05$, close to the median redshift of our SDSS catalog.
While SHAM in its simplest implementation assumes a one-to-one correspondence between $\mmax$ and $\mstar$, a scatter of $0.15 - 0.2$ dex in this relation is suggested by observations \citep[e.g.,][]{ZheCoiZeh07, YanMovdB08, MorvdBCac09, WetWhi10, LeaTinBun12}.
Thus, in our implementation we assume 0.15 dex log-normal scatter in $\mstar$ at fixed $\mmax$, achieved by deconvolving the observed SMF with a log-normal filter such that we recover the observed SMF after adding this scatter.

\subsection{Simulation group catalog} \label{sec:group_catalog_sim}

To make robust comparisons with our SDSS group catalog, we produce a `simulation group catalog' by applying the same group-finding algorithm that we use in SDSS to our simulation galaxy catalog.
(We use the distant observer approximation and do not produce a light-cone.)
While we base our models of SFR evolution on true satellite versus central demarcation in the simulation, we effectively `observe' the results at $z = 0.05$ through the simulation group catalog, which includes the effects of interloping galaxies caused by redshift-space distortions and any other systematics of the group-finding algorithm.
In \citetalias{WetTinCon13b}, we will show that the galaxy distributions in the simulation group catalog closely match those of the SDSS group catalogs.

Our simulation group catalog also allows us to correct for the effects from interloping galaxies in measuring satellite and central galaxy quiescent fractions in the SDSS group catalog.
Interlopers have little effect on central galaxy quiescent fractions because central galaxies strongly outnumber satellites, but interlopers do cause the observed satellite quiescent fractions to be $\sim 10\%$ too low (see Appendix C of \citetalias{TinWetCon11}).
To correct for these effects, we create a `mock' SDSS group catalog by empirically assigning SSFRs to galaxies in the simulation group catalog, matching the observed SSFR distribution separately for satellite and central galaxies in narrow bins of galaxy and host halo mass.
We then measure quiescent fractions in the mock catalog according to each galaxy's true (real-space) satellite/central designation, which provides values unbiased by redshift-space distortions.

Finally, our use of a simulation group catalog mitigates any inconsistency between the simulation's FoF halo definition, which allows arbitrary morphology, and the spherical overdensity halo definition applied to SDSS, because we always compare the two galaxy catalogs using the same halo definition.
For reference, $\mthm \approx 1.2\,M_{\rm FoF}(b=0.168)$.

\section{Satellite infall types and times} \label{sec:infall-time}

To inform our models for the evolution of satellite SFR and interpret our results, we first use the simulation to explore the pathways and times of infall for satellites at $z = 0.05$ (the median redshift of our SDSS catalog).
To highlight physical trends free from the ambiguities of redshift-space distortions, in this subsection we do not use the simulation group catalog, but we examine satellites based directly on the simulation halo catalog.

\subsection{Satellite infall and ejection} \label{sec:infall_eject}

\begin{figure}
\centering
\includegraphics[width = 0.99 \columnwidth]{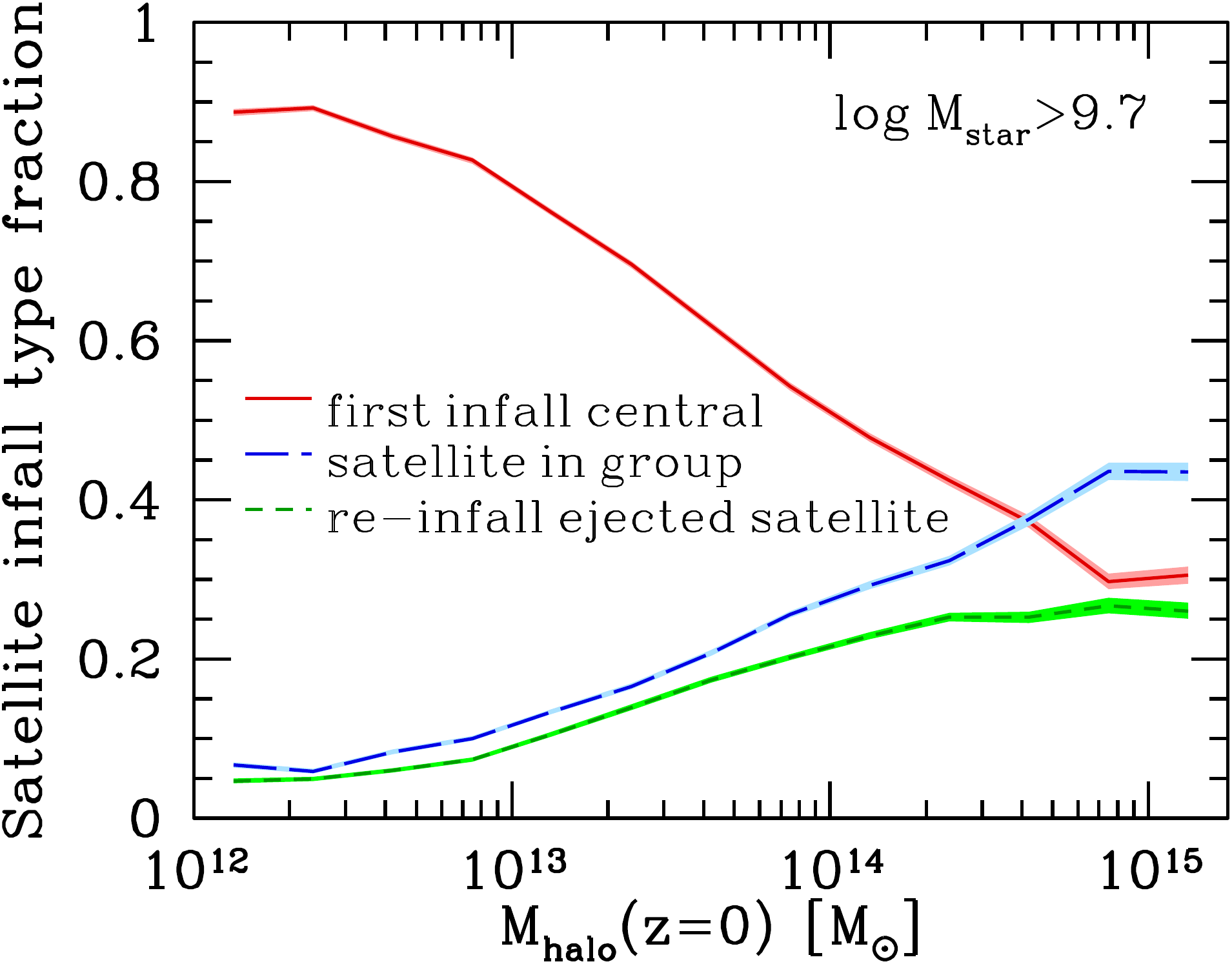}
\caption{
Infall type fractions for satellites at $z = 0.05$ in the simulation versus host halo mass.
Infall types are defined at the time of most recent infall into their current host halo.
Solid red curve shows satellites that fell in as a central galaxy, for the first time, directly from the field.
Short-dashed blue curve shows satellites that fell in as a satellite in a lower mass host halo (group infall).
Long-dashed green curve shows satellites that expirienced re-infall after being ejected beyond $\rvir$.
Shaded widths show beta distribution 68\% confidence intervals.
Satellite residing in higher mass host halos are more likely to have fallen in as a satellite or experienced an ejected phase.
The dependence on satellite stellar mass is much weaker (not shown, see text).
Fewer than half of satellites in halos $> 10 ^ {14} \msun$ fell in directly from the field.
} \label{fig:infall-type-frac_v_m-halo}
\end{figure}

In examining the infall times of satellites, we consider two ways to define infall: the time of most \tit{recent} infall into the \tit{current} host halo, or the time of \tit{first} infall into \tit{any} host halo.
The latter includes any time spent in a lower mass halo before falling into the current host halo.
The latter also naturally incorporates satellites that temporarily orbit beyond their host halo's virial radius, $\rvir$, which we call `ejected satellites', and then fall in again \citep{GilKneGib05, LudNavSpr09, WanMoJin09}.
If one considers only their most recent infall, these ejected satellites might appear to be central galaxies falling in directly from the field for the first time, but as we will show, it is more appropriate to consider them as satellites since their time of first infall.

We define the `first' infall of a subhalo as the first time that it becomes a satellite (is linked by the FoF algorithm) in another host halo, and `recent' infall as the time that it falls into the main progenitor of its current host halo.
In defining first infall, we additionally require that a subhalo remains a satellite for at least two consecutive simulation outputs to avoid cases of temporary bridging as a subhalo briefly passes through the outskirts of a host halo.

Fig.~\ref{fig:infall-type-frac_v_m-halo} shows the fraction of infall types for satellites in the simulation at $z = 0.05$, as a function of and current host halo mass.
Shaded widths show 68\% confidence interval as given by a beta distribution \citep{Cam11}.
In low-mass halos, 90\% of satellites in our mass range fell in as central galaxies directly from the field for the first time, but this fraction drops below 50\% in halos $> 10 ^ {14} \msun$.
Thus, \tit{within the most massive host halos, the dominant mode of infall is as a satellite in a lower mass halo} (group preprocessing).
So, group preprocessing of star formation is potentially important in this regime, as we will explore in \S\ref{sec:where_quench}.
We also examine the dependence on satellite stellar mass (not shown), though it is much weaker than for host halo mass, with the fraction that fell in as a satellite in another halo or for the second (or more) time after being ejected both dropping from 20\% at $\mstar = 10 ^ {9.7} \msun$ to 10\% at $\mstar = 10 ^ {11.3} \msun$.
These infall fractions agree broadly with those of previous works \citep{BerSteBul09, McGBalBow09, DeLWeiPog12}, though each of those works find quantitative differences arising largely from different satellite and group mass limits \citep[see discussion in][]{DeLWeiPog12}.

Fig.~\ref{fig:infall-type-frac_v_m-halo} also shows that satellites that have been ejected beyond their host halo's $\rvir$ play an important role in satellite evolution.
We find that these satellites spend typically $1 - 2 \gyr$ inside a host halo when they first fall in, experiencing a single pericentric passage before being ejected.
They then spend longer time ($2 - 3 \gyr$) orbiting as a central galaxy beyond $\rvir$ until they fall in again.
Both timescales have no significant dependence on satellite mass, but both do have broad distributions extending out to $4 - 5 \gyr$, implying that some satellites are kicked out via multi-body encounters after several orbital passages \citep{LudNavSpr09}.

After being ejected, $\gtrsim 90\%$ of these (now central) galaxies continue to lose halo mass, with a typical ejected satellite currently having half of the halo mass that it had at the time of ejection.\footnote{
If a subhalo grows in $\mmax$ by $> 50\%$ since ejection, we define it to be a `newly' formed halo and discard it from the ejected population, though this affects only a few percent of ejected satellites.}
This halo mass stripping occurs as ejected satellites orbit in the hot, dynamic environment surrounding a massive host halo.
This continued halo mass loss strongly suggests that ejected satellite galaxies evolve in a similar way as those within $\rvir$ and would also exhibit truncated SFRs.
In \citetalias{WetTinCon13a}, we will examine the trends of ejected satellites as a function of halo-centric distance, showing that the ejected fraction around massive halos can account for the enhanced quiescent fraction of central galaxies out to $\sim 2.5\,\rthm$ around massive clusters that we noted in \citetalias{WetTinCon12a} \citep[see also][]{WanYanMo09}.
Thus, we conclude that SFRs in ejected satellite continue evolve in a similar manner as those within $\rvir$, and we will treat the two populations identically in our models for SFR evolution.

\subsection{Satellite infall times} \label{sec:infall-time_v_m}

\begin{figure}
\centering
\includegraphics[width = 0.99 \columnwidth]{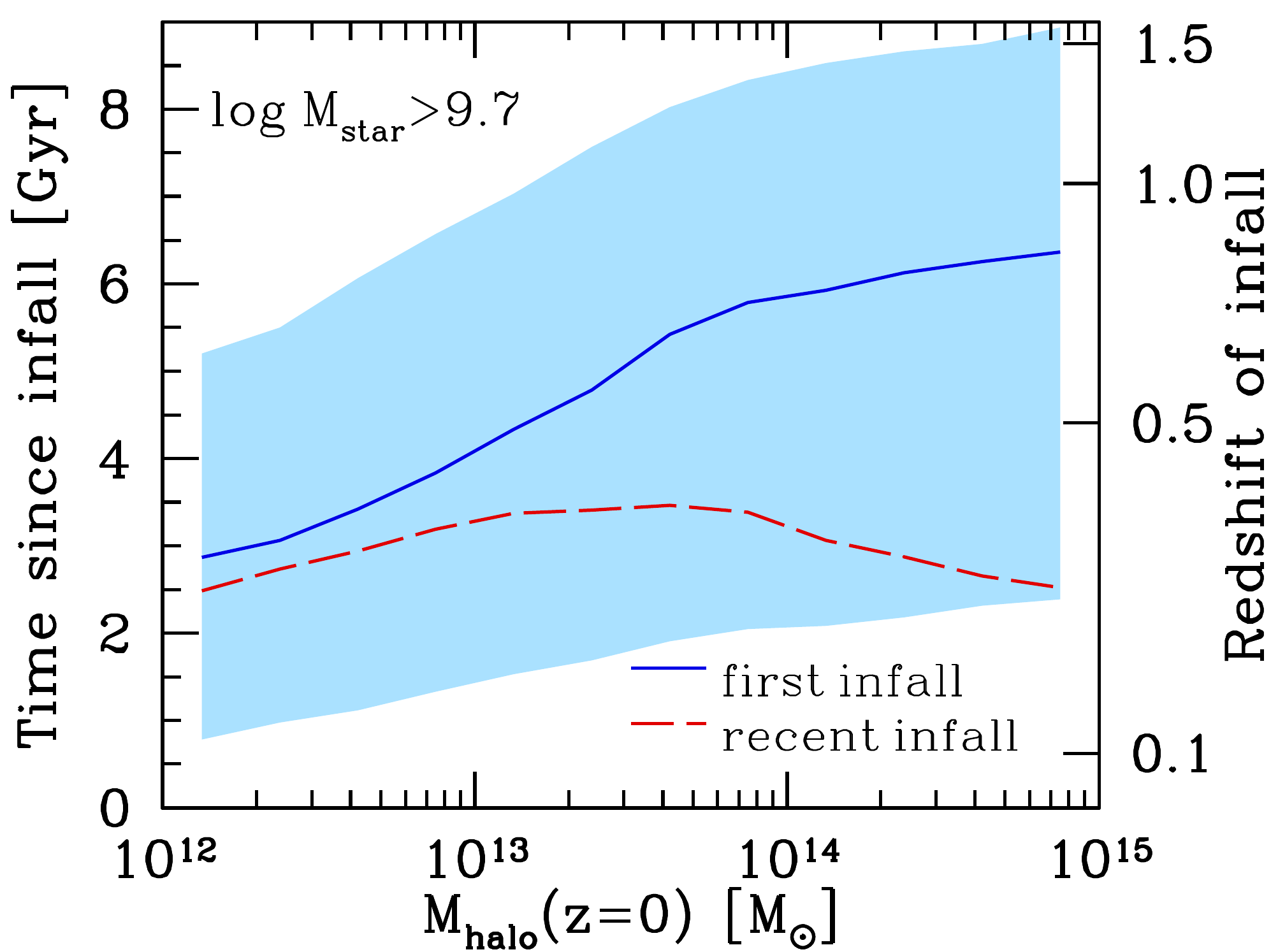}
\caption{
Median time since infall (or redshift of infall) versus host halo mass for all satellites in the simulation at $z = 0.05$.
Blue curve shows first infall into any host halo and red curve shows most recent infall into the current host halo.
Shaded region shows 68\% distribution interval, which is similar in magnitude for first and recent infall (latter not shown).
Time since recent infall exhibits no significant mass dependence.
Time since first infall exhibits a strong increase with host halo mass, driven by the increasing fraction of infalling groups and ejected satellites in more massive halos (Fig.~\ref{fig:infall-type-frac_v_m-halo}).
Time since first infall also exhibits a weak decrease with stellar mass (not shown, see text).
} \label{fig:t-inf_v_m-halo}
\end{figure}

With the above definitions of first and recent infall, Fig.~\ref{fig:t-inf_v_m-halo} shows how surviving satellites' time since infall (and redshift of infall) depends on their current host halo mass.
Time since \tit{recent} infall does not depend on either host halo mass or satellite stellar mass (latter not shown).
There is a slight decrease at high halo mass, because more massive halos formed more recently, gaining many satellites via the recent infall of groups (Fig.~\ref{fig:infall-type-frac_v_m-halo}).
By contrast, time since \tit{first} infall exhibits a strong increase with halo mass, a result of the increased fraction of infalling groups and ejected satellites from Fig.~\ref{fig:infall-type-frac_v_m-halo} \citep[see also][]{WanLiKau07, DeLWeiPog12}.
We also find that more massive satellites experienced their first infall slightly more recently, with median time since infall falling from $5 \gyr$ at $\mstar = 10 ^ {9.7} \msun$ to $3.5 \gyr$ at $\mstar = 10 ^ {11.3} \msun$.
This stellar mass dependence is caused by hierarchical structure growth, such that later infalling satellites had more time to grow in mass before infall, coupled with shorter lifetimes for more massive satellites, such that satellites that fell in early but have survived to the present are preferentially of lower mass.

Half of satellites in our mass range first fell in at $z \gtrsim 0.5$, with a broad tail out to $z \ge 1$ (shaded region), so they typically have experienced $\gtrsim 4 \gyr$ evolving as a satellite.
Furthermore, while a satellite typically has spent $\sim 3 \gyr$ in its current host halo, the timescale is twice as long if in a $> 10 ^ {14} \msun$ halo.
Thus, on average, satellites have spent 1/3 - 1/2 of their galactic lifetime evolving as a satellite, highlighting the importance of satellite evolution as a component of galaxy evolution.
These infall times also highlight the importance of obtaining accurate satellite SFR initial conditions prior to infall to understand their subsequent SFR evolution.

Finally, we emphasize that we limit our analysis to subhalos that our simulation resolves well (at least 1500 particles at infall), and that we have accounted for satellite merging/disruption in a way that matches various observed galaxy statistics, providing accurate satellite lifetimes.
Thus, we expect our results to be robust against resolution effects.

\subsection{Defining satellite infall}

We have shown that more than half of satellites in host halos $> 10 ^ {14} \msun$ fell in as a satellite in another halo or as an ejected satellite, and that, as a result, time since infall is higher in more massive halos if considering \tit{first} infall.
Combined with three key observational results from \citetalias{WetTinCon12a}, this provides strong evidence that the satellite-specific quenching process(es) begins at \tit{first} infall, regardless of the host halo's mass.
First, satellites exhibit an enhanced quiescent fraction even in the lowest mass host halos that we probed ($3 \times 10 ^ {11} \msun$).
To the extent that this remains true at higher redshift, then the satellite-specific process(es) begins upon infall into any halo, regardless its mass.
Second, central galaxies out to $\sim 2\,\rthm$ around massive host halos exhibit an enhanced quiescent fraction, implying that ejected satellites can become/remain quenched.
Third, the satellite quiescent fraction increases with host halos mass even at fixed projected distance from halo center, $\dproj / \rvir$, including satellites at $\dproj \approx \rvir$, which tend to have fallen into their \tit{current} halo recently.
More specifically, at fixed $\dproj / \rvir$, time since \tit{recent} infall does not increase with halo mass, but time since \tit{first} infall does (see \citetalias{WetTinCon13b}).

Taken together, these constitute strong evidence that satellite-specific quenching process(es) starts to set in at \tit{first} infall, and we will use only this definition of infall henceforth.
For a satellite that experienced first infall at time $\tinf$, we refer to its time since first infall, $\tsinf = t - \tinf$.

\section{The quenching of star formation in satellites}

In this section, we present our main results on satellite quenching.
First, in \S\ref{sec:sfr_at_infall}, we construct an empirical parametrization for the evolution of SFRs of central galaxies to provide accurate initial conditions for the SFRs of satellites at their time of first infall.
Then, in \S\ref{sec:quench_importance_efficiency} we explore the importance and efficiency of satellite quenching in an empirical manner by combining satellite infall times with our parametrization for initial quiescent fractions from \S\ref{sec:qu.frac-evol_cen}.
Finally, in \S\ref{sec:sfr-evol_sat} we develop models for satellite SFR evolution after infall to constrain the timescales over which satellites are quenched.

\subsection{SFR in satellites at the time of first infall} \label{sec:sfr_at_infall}

To understand the evolution of SFR in satellites after infall, we first need accurate initial conditions for their SFRs just prior to infall, as given by the SFRs of central galaxies of the appropriate stellar mass at a satellite's redshift of infall.
As we showed in \S\ref{sec:infall-time_v_m}, satellites in our mass range first fell in typically at $z \approx 0.5$, with a broad distribution out to $z > 1$, so using $z \approx 0$ central galaxy properties as the initial conditions for satellites, which has been a common approach \citep[e.g.,][]{vdBAquYan08, TinWet10, PenLilRen11, DeLWeiPog12}, is a poor approximation.

Thus, we require a statistical parametrization of the full SFR distribution of central galaxies as a function of both stellar mass and redshift.
In order to describe this quantity both accurately and in a manner that is independent of (and minimally degenerate with) our models of satellite SFR evolution after infall, we proceed in an empirical manner.
While some previous approaches have modeled the evolution of galaxy stellar mass and SFR empirically \citep[e.g.,][]{ConWec09}, these studies examined only the \tit{average} SFRs of galaxies, not taking into account the bimodal nature of the SFR distribution, in particular, quiescent galaxies.
Our approach is more comprehensive, as our evolution parametrization contains two key components: the fraction of central galaxies that are quiescent and the normalization of galaxy SFRs.
We describe these in turn.

\subsubsection{Evolution of the quiescent fraction for central galaxies} \label{sec:qu.frac-evol_cen}

Our goal is to start with observations of the evolution of the quiescent fraction for \tit{all} galaxies out to $z = 1$ and decompose this into the evolution for satellite and central galaxies separately, thus allowing us to use the values for central galaxies to provide the initial quiescent fractions for satellites at infall.
In order to quantify reasonable systematic modeling uncertainty, we implement two contrasting parameterizations for this satellite-central decomposition at $z > 0$.
Our `fiducial' parametrization separates satellite and central galaxies based on analysis of spatial clustering measurements \citep{TinWet10}: here, the satellite quiescent fraction does not evolve at fixed stellar mass, and the central galaxy quiescent fraction declines rapidly with redshift.
As an `alternative' parametrization, we allow both satellite and central galaxy quiescent fractions to evolve at the same rate, which leads to the central galaxy quiescent fraction declining more gradually with redshift.
We include both of these contrasting parametrizations with the goal of bracketing reasonable modeling uncertainty in our approach.

We begin by describing our fiducial parametrization.
We can obtain the fraction of central galaxies that are quiescent, $\fcenq = \ncenq / \ncen$, by knowing the fraction of all galaxies that are quiescent, $\fallq = \nallq / \nall$, the fraction of all galaxies that are satellites, $\fsat = \nsat / \nall$, and the fraction of satellites that are quiescent, $\fsatq = \nsatq / \nsat$, via
\begin{equation} \label{eq:qu.frac_v_z_cen}
\fcenq = \frac{\fallq - \fsatq \fsat}{1 - \fsat} \, ,
\end{equation}
in which each fraction depends on stellar mass and redshift.

\begin{table*}
\centering
\begin{tabular}{|c|c|cccc|}
\hline
\multicolumn{6}{c}{Fiducial parametrization} \\
\multicolumn{6}{c}{$\fcenq(\mstar, z) = \frac{\fallq(\mstar, z) - \fsatq(\mstar) \fsat(\mstar, z)}{1 - \fsat(\mstar, z)}$} \\
\hline
Fit & Parameter & \multicolumn{4}{c}{Value} \\
& & \multicolumn{4}{c}{$\log(\mstar / \msun)$}\\
& & $9.5-10.0$ & $10.0-10.5$ & $10.5-11.0$ & $11.0-11.5$ \\
\hline
\multirow{2}{*}{$\fallq(z) = A (1 + z)^\alpha$} & $A$ & 0.227 & 0.471 & 0.775 & 0.957 \\
& $\alpha$ & -2.1 & -2.2 & -2.0 & -1.3 \\
\multirow{2}{*}{$\fsat(z) = B_0 + B_1 z$} & $B_0$ & 0.33 & 0.30 & 0.25 & 0.17 \\
& $B_1$ & -0.055 & -0.073 & -0.11 & -0.10 \\
\multirow{2}{*}{$\fsatq(\mstar) = C_0 + C_1 \log(\mstar)$} & $C_0$ & \multicolumn{4}{c}{-3.26} \\
& $C_1$ & \multicolumn{4}{c}{0.38} \\
\hline

\hline
\multicolumn{6}{c}{Alternate parametrization} \\
\multicolumn{6}{c}{$\fcenq(\mstar, z) = \fcenq(\mstar, z = 0) \times (1 + z) ^ \alpha$} \\
\hline
\multirow{2}{*}{$\fcenq(\mstar, z = 0) = D_0 + D_1 \log(\mstar)$} & $D_0$ & \multicolumn{4}{c}{-6.04} \\
& $D_1$ & \multicolumn{4}{c}{0.63} \\
\hline
\end{tabular}
\caption{
Fits of the quiescent fraction for central galaxies, as a function of stellar mass and redshift.
\textbf{Top}: Fiducial parametrization, based on equation (\ref{eq:qu.frac_v_z_cen}), which incorporates the quiescent fraction for all galaxies, $\fallq(z)$, the satellite fraction, $\fsat(z)$, and the quiescent fraction for satellites, $\fsatq(\mstar)$.
Fits to $\fallq(z)$ and $\fsat(z)$ are given in stellar mass bins, while $\fsatq(\mstar)$ is fit to stellar mass dependence assuming no redshift dependence.
\textbf{Bottom}: Alternate parametrization, based on equation (\ref{eq:qu.frac_v_z_cen_alt}).
$\fcenq(\mstar)$ is fit to stellar mass dependence at $z = 0$, and $\alpha$, in bins of stellar mass, is the same as in the fiducial parametrization.
} \label{tab:qu.frac_v_z_cen}
\end{table*}

\begin{figure}
\centering
\includegraphics[width = 0.99 \columnwidth]{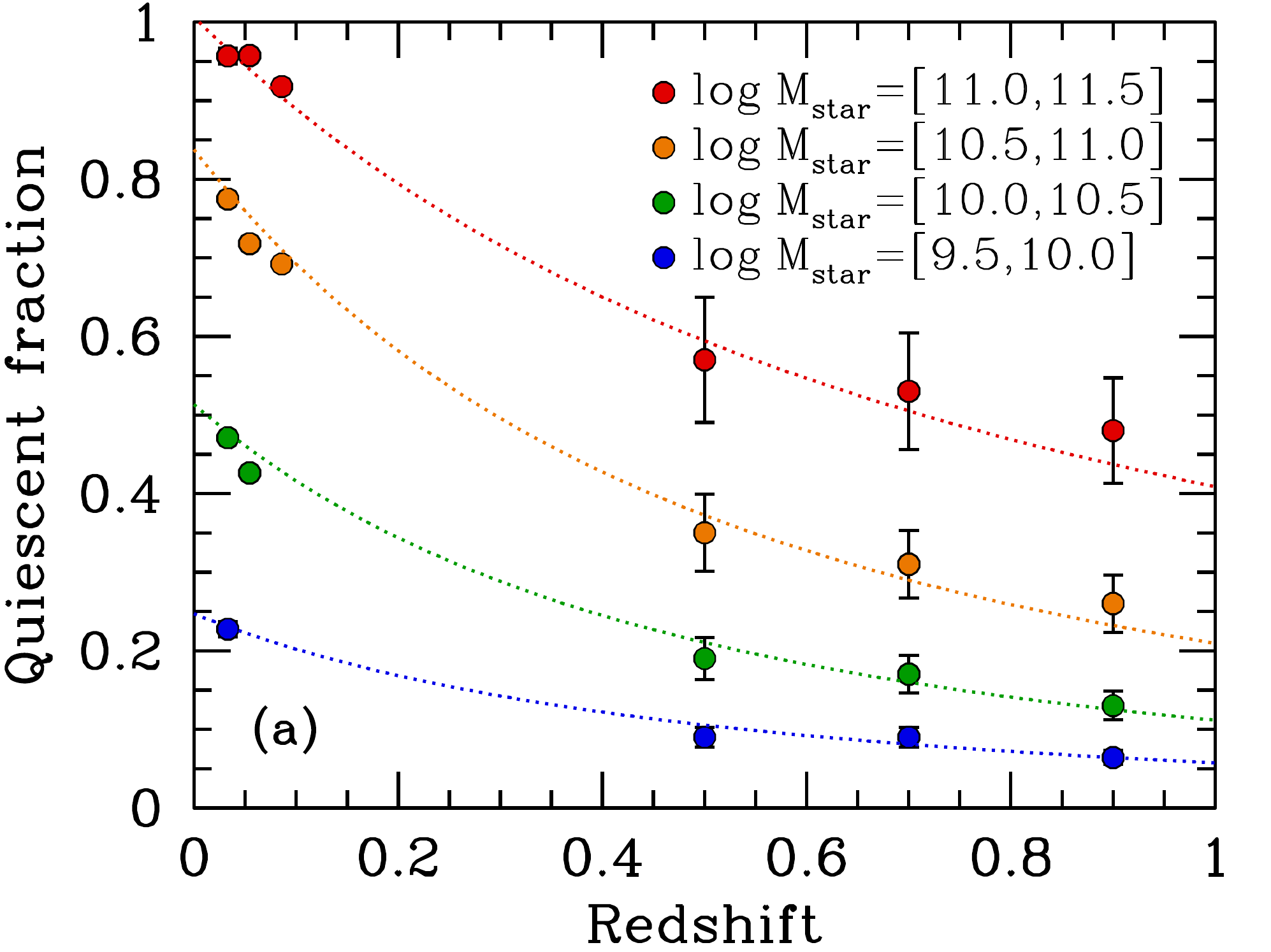}
\includegraphics[width = 0.99 \columnwidth]{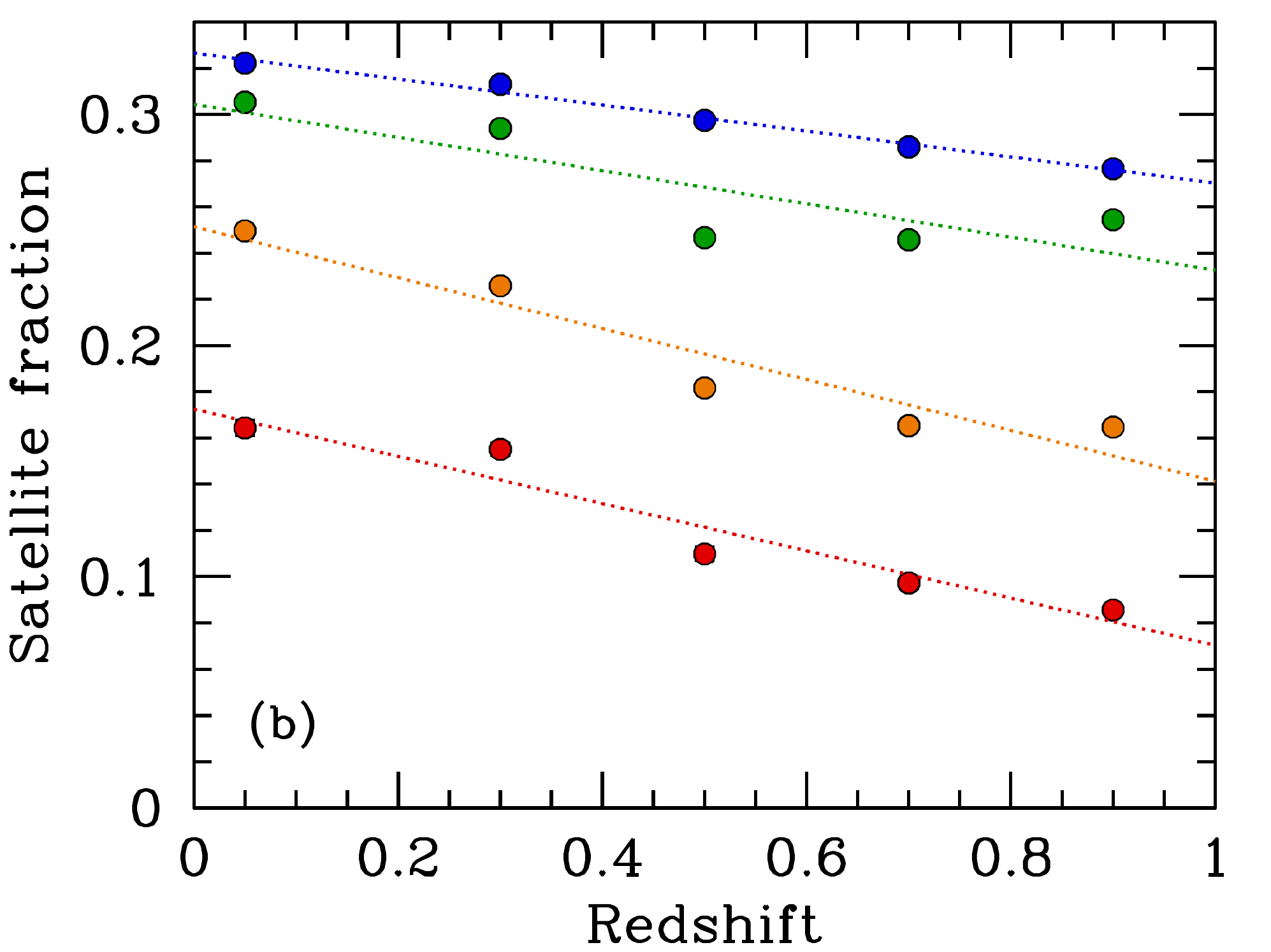}
\includegraphics[width = 0.99 \columnwidth]{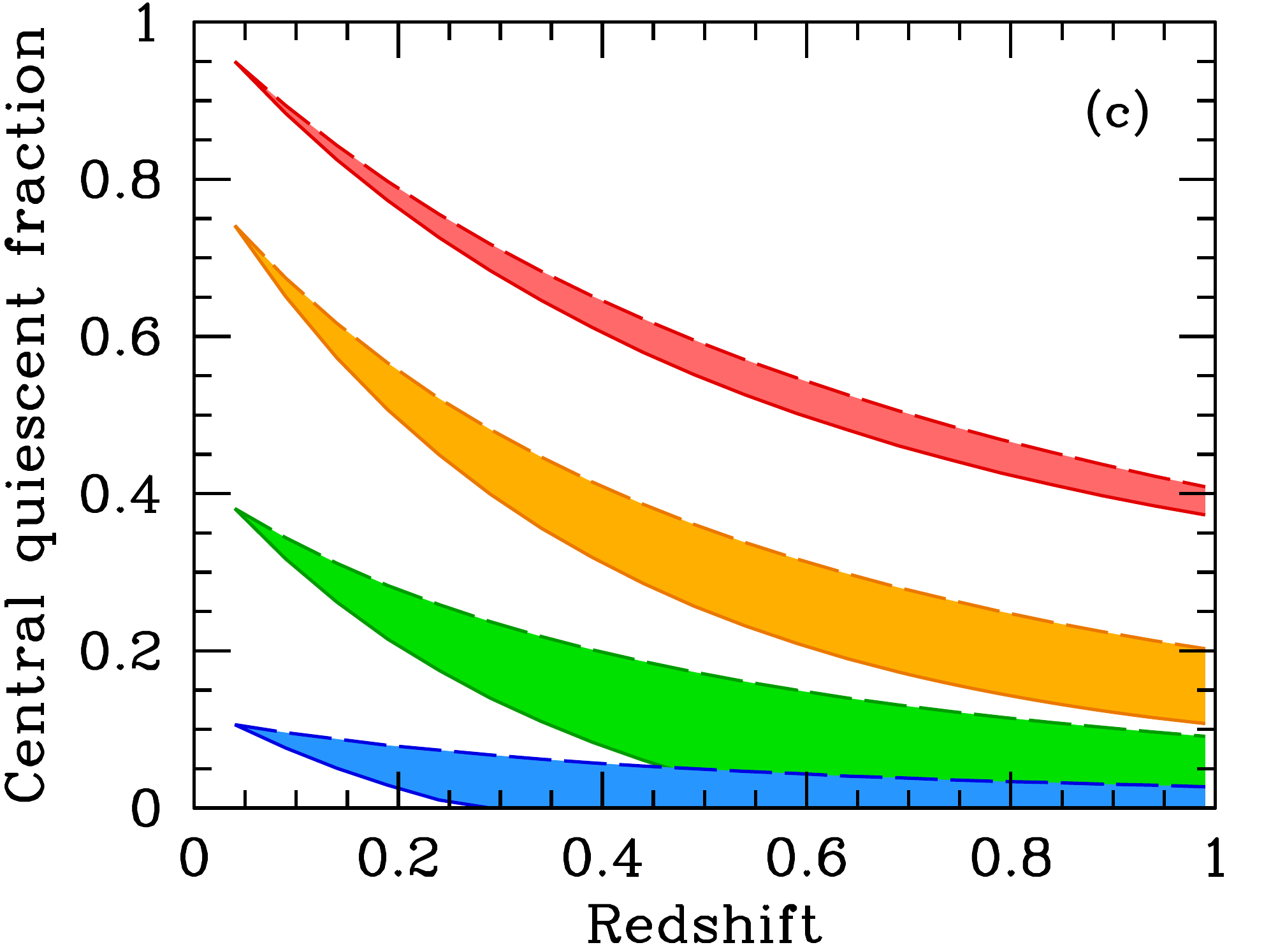}
\caption{
\textbf{(a)}: Fraction of all galaxies that are quiescent versus redshift, in bins of stellar mass.
Points at $z \approx 0$ show our results from SDSS, while points at higher redshift are from the Cosmic Evolution Survey (COSMOS) \citep{DroBunLea09}.
Curves show best-fit evolution in each mass bin.
\textbf{(b)}: Fraction of galaxies that are satellites versus redshift from the simulation, using SHAM to assign stellar mass at each redshift.
Curves show best-fit evolution in each mass bin.
\textbf{(c)}: Fraction of central galaxies that are quiescent versus redshift.
Solid curves show lower limits from equation (\ref{eq:qu.frac_v_z_cen}) and dashed curves show upper limits from equation (\ref{eq:qu.frac_v_z_cen_alt}).
The difference from (a) is small at high mass but more significant at low mass.
Note that the quiescent fractions in (a) and (c) monotonically increase with stellar mass, while the satellite fractions in (b) monotonically decrease with stellar mass.
} \label{fig:qu.frac_v_z}
\end{figure}

\begin{figure}
\centering
\includegraphics[width = 0.99 \columnwidth]{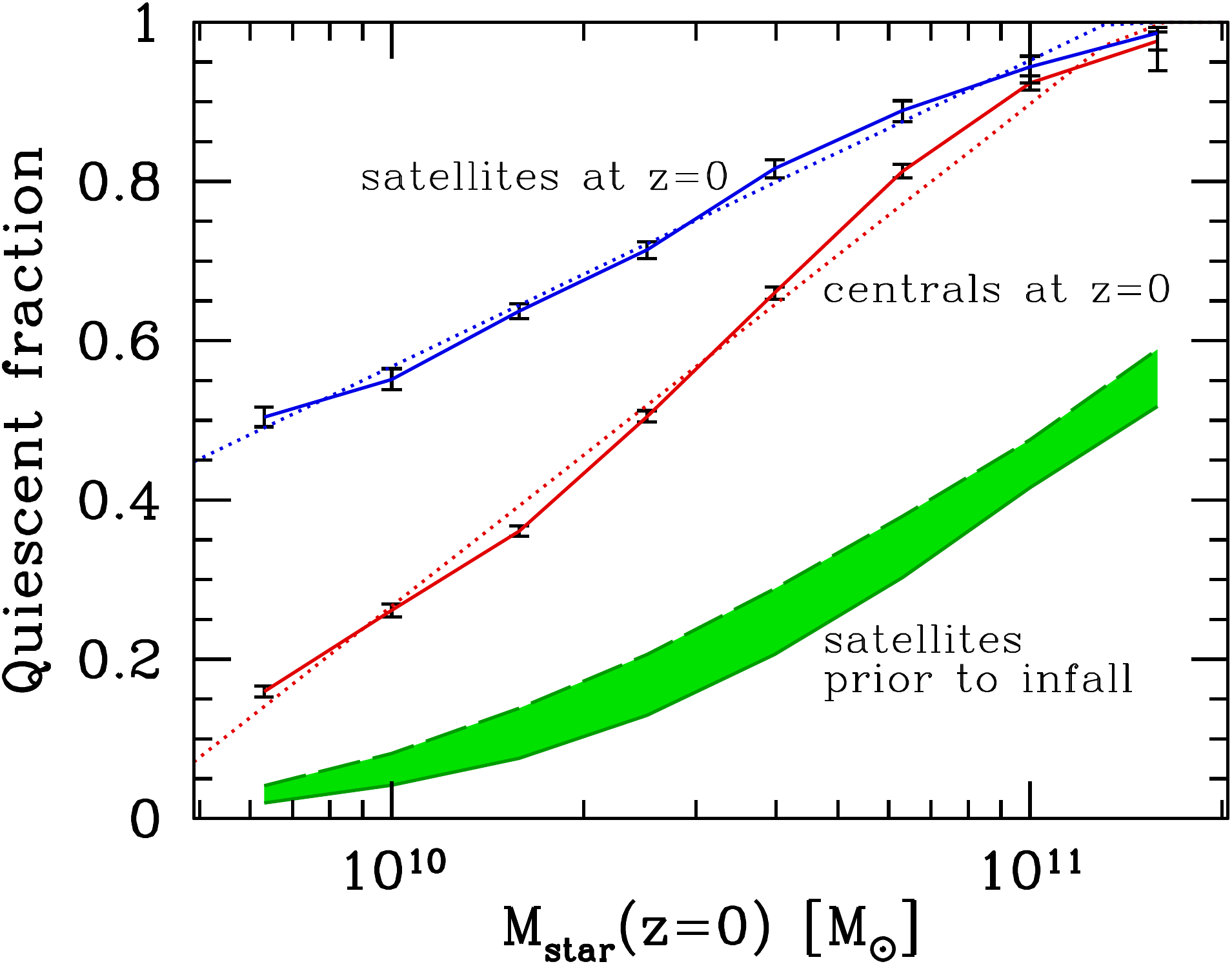}
\caption{
Fraction of galaxies that are (were) quiescent versus stellar mass.
Solid curves show values at $z = 0$ measured in the SDSS group catalog (using the simulation group catalog to correct for interloping galaxies), for satellite (blue, top) and central (red, middle) galaxies.
Dotted lines show fits, given in Table~\ref{tab:qu.frac_v_z_cen}.
Green region at bottom shows what fraction of satellites at $z = 0$ were quiescent prior to their first infall, with uncertainty given by our two parametrizations.
The significant increase of satellites' quiescent fraction from first infall (green) to $z = 0$ (blue) highlights the importance of satellite quenching.
The difference between the quiescent fraction for satellites prior to first infall (green) and for central galaxies at $z = 0$ (red) shows that the latter does not represent accurate initial conditions for satellites.
} \label{fig:qu.frac_v_m-star}
\end{figure}

To determine the quiescent fraction for all galaxies, $\fallq(\mstar, z)$, we combine our SDSS results for all galaxies at $z = 0$ with quiescent fractions from the COSMOS survey at $z < 1$ \citep{DroBunLea09}.
While the results of \citeauthor{DroBunLea09} are based on photometric spectral energy distributions (SED), the significant number of photometric bands (30) in COSMOS helps to ensure accurate redshifts, stellar masses, and active versus quiescent demarcations.
In particular, \citeauthor{DroBunLea09} identified active versus quiescent galaxies using full SED fitting, which minimizes the effects of dust contamination as compared with simpler color cuts.\footnote{
Despite using the same IMF, the SED-based stellar masses in \citeauthor{DroBunLea09} are estimated to be $\sim 0.2$ dex higher on average than those in our SDSS catalog.
However, their stellar masses also have larger scatter because of their photometric redshifts.
In terms of quiescent fractions, these effects largely cancel out, so we do not attempt to renormalize stellar masses.
See \citetalias{Tin12}.}
Fig.~\ref{fig:qu.frac_v_z}a shows the evolution of the quiescent fraction in bins of stellar mass, along with the best-fit relation in each mass bin
\begin{equation} \label{eq:qu.frac_v_z}
\fallq(\mstar, z) = A(\mstar) \times (1 + z) ^ {\alpha(\mstar)} \,.
\end{equation}
We split our SDSS sample into narrow redshift bins, and we anchor the fit to the quiescent fraction in the lowest ($z < 0.04$) bin.
Note that the evolution within SDSS broadly agrees with the fits to much higher redshift.
In all mass bins, the quiescent fraction has at least doubled since $z = 1$.

To determine the satellite fraction, $\fsat(\mstar, z)$, we use the simulation directly, motivated by the agreement (within observational uncertainty) of this quantity between the simulation and our SDSS group catalog at $z = 0$ \citep[see also][]{WetWhi10}.
We use SHAM to assign stellar mass to subhalos at each redshift, using the SMF from SDSS \citep{LiWhi09} at $z = 0.1$ and from COSMOS \citep{DroBunLea09} at $z = 0.3, 0.5, 0.7, 0.9$, assuming 0.15 dex $\mstar - \mmax$ scatter in all cases.
Fig.~\ref{fig:qu.frac_v_z}b shows the evolution of the satellite fraction in stellar mass bins.
We find that linear growth with redshift at fixed mass, as given by
\begin{equation} \label{eq:sat.frac_v_z}
\fsat(\mstar, z) = B_0(\mstar) + B_1(\mstar) z \, ,
\end{equation}
provides a reasonable fit, as Fig.~\ref{fig:qu.frac_v_z}b shows.
Note that the fluctuations with redshift are driven by the evolution of the SMF in \citeauthor{DroBunLea09} and not by subhalo statistics.

Finally, to determine the quiescent fraction for satellites, $\fsatq(\mstar, z)$, we first fit to the stellar mass dependence at $z = 0$ from our SDSS group catalog, as Fig.~\ref{fig:qu.frac_v_m-star} shows.
Recall that we use the simulation group catalog to remove the effects of galaxy interlopers caused by redshift-space distortions.
We find that
\begin{equation} \label{qu.frac_v_m_sat}
\fsatq(\mstar, z = 0) = C_0 + C_1 \log(\mstar)
\end{equation}
provides a reasonable fit in our mass range, as shown by the dotted curve.
We then impose that there is no evolution of this quantity, such that $\fsatq(\mstar, z) = \fsatq(\mstar, z = 0)$.
This choice is motivated by the results of \citet{TinWet10}, who found no evolution in the quiescent fraction for satellites at fixed magnitude at $z \le 1$, based on halo occupation modeling of the spatial clustering and number densities of galaxy samples from the Classifying Objects by Medium-Band Observations (COMBO-17) \citep{PhePeaMei06} and Deep Extragalactic Evolutionary Probe (DEEP2) \citep{CoiNewCro08} surveys.
Note that \citetalias{Tin12} find similar results from the spatial clustering of COSMOS galaxies.

Putting these ingredients into equation (\ref{eq:qu.frac_v_z_cen}), we obtain the quiescent fraction for central galaxies as a function of stellar mass and redshift, as Fig.~\ref{fig:qu.frac_v_z}c shows (solid curves).
The difference from the overall quiescent fraction (panel a) is modest at high mass but is more significant at low mass, where central versus satellites quiescent fractions differ more strongly and the satellite fraction is higher.

This fiducial parametrization is based on the satellite quiescent fraction not evolving at fixed stellar mass since $z = 1$.
This behavior is motivated by spatial clustering measurements and also is supported by \citet{GeoLeaBun11}, who examined galaxy groups of mass $10 ^ {13 - 14} \msun$ in COSMOS out to $z = 1$ and found no significant evolution in the quiescent fraction of group members, at least for sufficiently massive galaxies ($\mstar > 3 \times 10 ^ {10} \msun$) that their sample is complete.
However, several other works observe that the quiescent/red fraction of galaxies in clusters decreases with increasing redshift \citep[e.g.,][]{ButOem84, PogvdLDeL06, McGBalWil11}.
In order to investigate this possible systematic uncertainty, we develop an alternate parametrization to quantify the impact of non-zero evolution of the satellite quiescent fraction at fixed stellar mass.
Motivated by the idea that, at least out to $z \sim 1$, the quiescent fraction for satellites is always higher than for central galaxies of the same mass, as supported by observations \citep[e.g.,][]{CooNewCoi07, McGBalWil11, GeoLeaBun11}, we parametrize the quiescent fractions for both satellite and central galaxies as decreasing with redshift at the same rate as for all galaxies, such that $\fcenq / \fsatq$ remains fixed.\footnote{
In detail, $\fcenq / \fsatq$ must evolve somewhat because $\fsat$ evolves, but $\fsat$ evolution causes $\fcenq / \fsatq$ to decrease by $< 10\%$ to $z = 1$.}
That is, at fixed mass, central galaxies evolve as
\begin{equation} \label{eq:qu.frac_v_z_cen_alt}
\fcenq(\mstar, z) = \fcenq(\mstar, z = 0) \times (1 + z) ^ {\alpha(\mstar)} \, ,
\end{equation}
using the same value of $\alpha$ as for all galaxies in the same stellar mass bin.
We fit for \fcenq(\mstar, z = 0) directly from the group catalog, as Fig.~\ref{fig:qu.frac_v_m-star} shows, finding that
\begin{equation} \label{eq:qu.frac_v_m_cen_alt}
\fcenq(\mstar, z = 0) = D_0 + D_1 \log(\mstar / \msun)
\end{equation}
provides a reasonable fit.
The dashed curves in Fig.~\ref{fig:qu.frac_v_z}c show the resultant evolution of the central galaxy quiescent fraction in this alternate parametrization, which exhibits a more gradual decline with redshift.
However, for both parametrizations, \tit{the central galaxy quiescent fraction increases by at least a factor of two from $z = 1$ to 0}.

Again, we emphasize that our two parametrizations provide significant contrast, and we consider their difference as representative of the reasonable systematic uncertainty in satellite initial conditions.\footnote{
We also tried an extreme scenario in which satellites are \tit{solely} responsible for the evolution of the quiescent fraction for all galaxies, to see if it is possible that the quiescent fraction for central galaxies does not evolve at fixed mass.
However, even in the extreme scenario of no quiescent satellites by $z = 1$, the central galaxy quiescent fraction at fixed mass still \tit{must} decrease by a factor of at least 2 from $z = 0$ to 1 to account for the overall galaxy quiescent fraction --- there are simply not enough satellites.}
Table~\ref{tab:qu.frac_v_z_cen} lists the fits and parameters for each term in equations (\ref{eq:qu.frac_v_z_cen}) - (\ref{eq:qu.frac_v_m_cen_alt}).
For binned parameters, we will use spline interpolation as a function of $\log\left(\mstar\right)$ to obtain smooth stellar mass dependence.

\subsubsection{Evolution of SFR for central galaxies} \label{sec:sfr-evol_cen}

Having parametrized the evolution of the quiescent fraction for central galaxies, we now develop a prescription for the evolution of their SFRs, which are observed to increase with redshift \citep[e.g.,][]{NoeWeiFab07}.
For central galaxies that remain active at $z = 0$, we parametrize their star formation history through a modified exponential $\tau$ model,
\begin{eqnarray} \label{eq:sfr-evol_cen}
\sfrcen(t) & \propto & (t - t_f) \exp \left\{ -\frac{(t - t_f)}{\taucen} \right\} \\
\mstar(t) & = & f_{\rm retain} \int_{t_f} ^ t \sfr(t) \, \rm{d}t \notag
\end{eqnarray}
with $t_f$ being the time of initial formation, which we take to be $t(z = 3)$ for all galaxies in our mass range, and $f_{\rm retain}$ being the fraction of stellar mass that is not lost through supernovae and stellar winds, which we take to be $f_{\rm retain} = 0.6$.\footnote{
In our model, this mass loss occurs instantaneously, though in reality, it is an extended process.
As \citet{LeiKra11} showed, the vast majority ($\sim 90\%$) of stellar mass loss occurs within the first $2 \gyr$, so our approximation is good given that 90\% of satellites fell in earlier than $2 \gyr$ ago.}
To obtain $\taucen$, we place all active central galaxies in our SDSS group catalog into narrow (0.2 dex) bins of stellar mass and compute the median $\taucen$ in each bin using equation (\ref{eq:sfr-evol_cen}) with the stellar mass and median SSFR of the bin.
This yields $\taucen$ values that range from 3.8 to $1.9 \gyr$ from $\mstar = 5 \times 10 ^ {9}$ to $2 \times 10 ^ {11} \msun$.
We use the median $\taucen$ of each bin to evolve back the SFRs of all active galaxies in that bin, which agree with measured SFRs at $0.2 < z < 1.1$ from \citet{NoeWeiFab07} to within 0.15 dex across our mass range, well within their measurement errors.
This prescription keeps the width of the active galaxy SFR distribution fixed, also in agreement with \citeauthor{NoeWeiFab07}.
Thus, by design, our $\tau$ model agrees well with the full active galaxy SFR distributions from \citeauthor{NoeFabWei07} out to $z = 1$.
While our $\tau$ model is highly simplified compared with the actual star formation histories of galaxies, we note that similar $\tau$ models have been shown to agree with observed average SFRs out to $z = 1$ (e.g., \citealt{NoeFabWei07}b).
Thus, for our specific purpose of assigning statistically accurate SFRs to active satellites at their time of infall, we consider this simplified but constrained model a reasonable empirical approach.

Parametrizing the possible change in the SFRs of quiescent central galaxies is more difficult, given both noisier measurements at $z = 0$ and a lack of detailed SFR measurements at higher redshift.
However, these uncertainties are largely irrelevant for us in practice, because satellites that fall in being already quiescent are at or quickly evolve below $\ssfr \approx 10 ^ {-12} \yrinv$, where our SDSS measurements are largely upper limits.
For completeness, in our parametrization we assume that quiescent galaxies at all redshifts have the same SFR normalization as quiescent galaxies at $z = 0$.
We also explored letting the SFR normalization for quiescent galaxies evolve by the same amount as for active galaxies, such that the separation between the peaks in the SFR distribution remains unchanged, though doing this does not alter significantly our model results at $z = 0$ in \S\ref{sec:quench_times}.

Note that equation (\ref{eq:sfr-evol_cen}) ignores the contribution from galaxy mergers when computing stellar mass growth, which means that it underestimates the amount of total stellar mass growth since $z = 1$ somewhat.
We examined the importance of this effect by using the stellar masses of galaxies from SHAM out to $z = 1$ and following galaxy-galaxy merger histories in our simulation to $z = 0$.
The typical amount of stellar mass growth via mergers since $z = 1$ for galaxies in our stellar mass range is always $< 10\%$ (usually much less); mergers become important only for galaxies of significantly higher mass.
Thus, this introduces only a small bias in our parametrization as compared with modeling uncertainty of initial quiescent fractions in \S\ref{sec:qu.frac-evol_cen}.

\begin{figure}
\centering
\includegraphics[width = 0.99 \columnwidth]{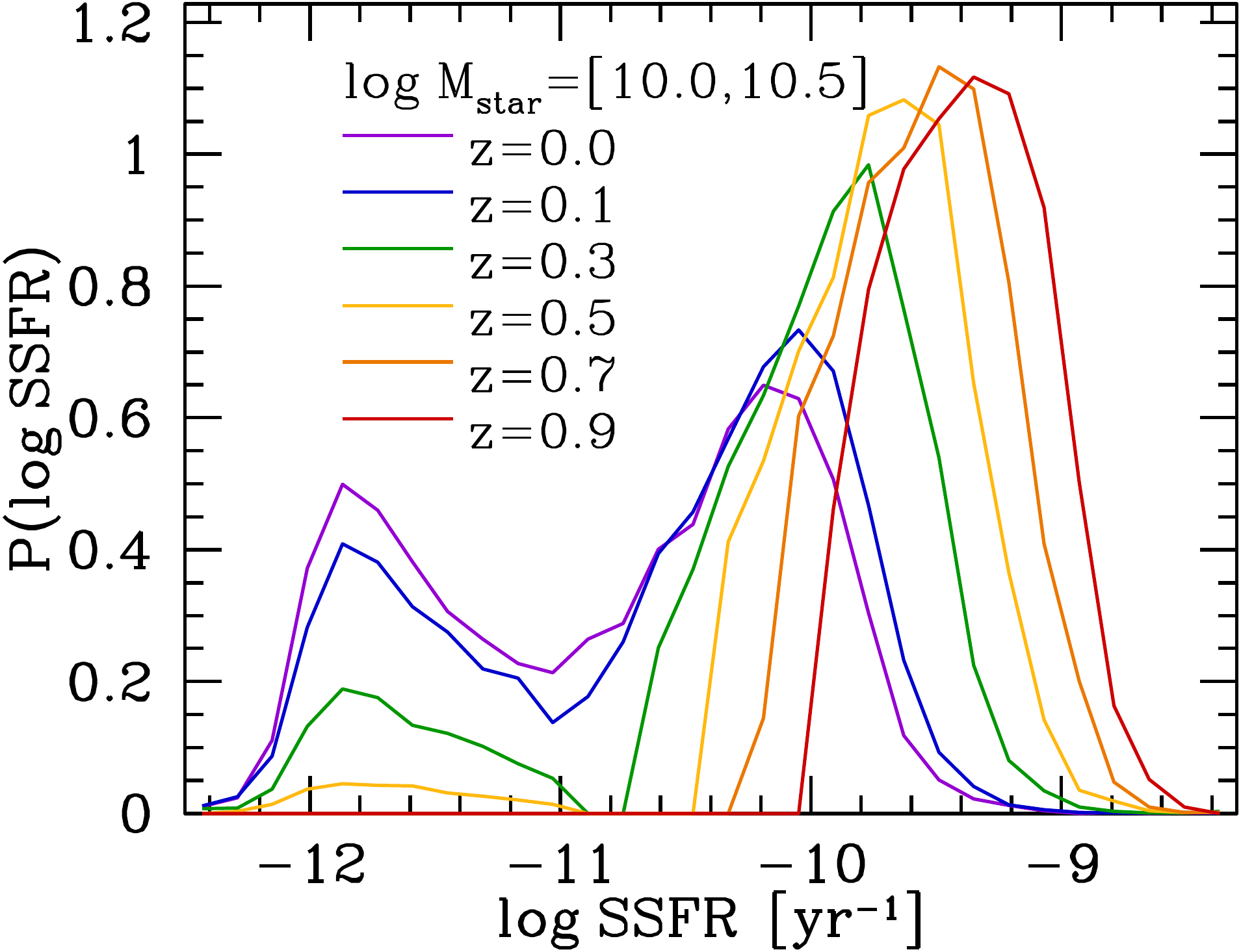}
\caption{
Example of the evolution of the SSFR distribution for central galaxies at fixed stellar mass.
The relative fraction of active and quiescent galaxies at each redshift is given by Table~ \ref{tab:qu.frac_v_z_cen} (fiducial parametrization), and the SSFR normalization of active galaxies increases with redshift according equation (\ref{eq:sfr-evol_cen}), using the median $\taucen$ of active central galaxies in the mass range.
} \label{fig:ssfr-distr_v_z}
\end{figure}

In summary, to obtain the SFR distribution of central galaxies as a function of mass and redshift we first compute the median $\taucen$ for active central galaxies at a given stellar mass, we use this $\taucen$ to compute the increase in SFR to a given redshift, and we apply this increase to all active central galaxies at that mass, such that the active galaxy SFR distribution width does not evolve at fixed mass.
We also assume that the SFR normalization for quiescent central galaxies does not evolve.
We then evolve the relative fraction of active and quiescent central galaxies according to equation (\ref{eq:qu.frac_v_z_cen}) or (\ref{eq:qu.frac_v_z_cen_alt}) by moving randomly chosen SFR values from the quiescent to the active side of the distribution.

Fig.~\ref{fig:ssfr-distr_v_z} shows an example of the resultant evolution of the SSFR distribution, for central galaxies with $\mstar = 10 ^ {10 - 10.5} \msun$.
We emphasize that our methodology reproduces the observed evolution of the quiescent fraction as well as the active galaxy SFR normalization and distribution width, providing accurate (statistical) initial conditions for the SFRs of satellites prior to first infall.
We also note that, while our parametrization clearly also has interesting implications for the physics of how central galaxies evolve, we defer further investigation in this area to future work.
For this paper, our parametrization is important only in providing accurate initial conditions for satellite SFRs.

\subsubsection{Assigning SFR to satellites at infall} \label{sec:assign_sfr_at_infall}

We now describe how we assign initial SFRs to satellites at their time of first infall.
For each satellite in the simulation at $z = 0$, we have its time of first infall, but in order to use our above parametrization, we must also know what stellar mass each satellite at $z = 0$ had at that time, at least in a statistically average sense.
To estimate this, we posit that satellites have grown in stellar mass by the same amount, on average, as active central galaxies of the same stellar mass.
This ansatz allows us to use equation (\ref{eq:sfr-evol_cen}), with the same median value of $\taucen$ as for central galaxies of the same stellar mass at $z = 0$, to calculate the average factor by which satellites were less massive at their time of first infall.
As we will demonstrate in \S\ref{sec:m-star_growth}, this ansatz is not only self-consistent, but moreover it \tit{must} be satisfied in our parametrization, at least in the absence of significant, systematic stellar mass loss from tidal stripping.\footnote{
If surviving satellites have experienced significant, systematic stellar mass loss from tidal stripping, this would mean that they were more massive at infall, so their initial quiescent fractions were higher, than in our model.
However, for feasible amounts of average stripping of $\lesssim 30\%$ (see Appendix \ref{sec:mass_growth_sham}), Fig.~\ref{fig:qu.frac_v_m-star} (green region) shows that the initial quiescent fractions would not increase by more than $\sim 5\%$.}

Note that this approach, based on statistical averaged star formation histories, ignores scatter in stellar mass growth.
As a check on the accuracy and self-consistency of our approach, we also tried computing stellar mass growth for central galaxies directly from SHAM, by differencing each central galaxy's stellar mass at $z = 0$ from what it had at a given redshift according to SHAM.
While this alternate approach leads to higher scatter in the amount of stellar mass growth over a given time interval (for example, $20 - 30\%$ since $z = 1$), it leads to average stellar mass growths that are consistent with the above method to within 10\%.
Thus, our approach is self-consistent, at least in parametrizing the average stellar mass growth of active galaxies.

Thus, for each satellite (including ejected satellites) in the simulation at $z = 0$, we assign its SFR at its redshift of first infall by drawing randomly from the central galaxy SFR distribution at the appropriate stellar mass at that redshift, using equation (\ref{eq:sfr-evol_cen}).
For satellites that fell in prior to $z = 1$, we extrapolate the quiescent fraction and $\tau$ model fits to higher redshift, though most satellites that fell in at $z > 1$ are low-mass and had minimal likelihoods of being quiescent at infall, so changing the extrapolation changes our results by only a few percent.

To highlight the importance of our parametrization for assigning accurate initial quiescent fractions to satellites, the green region in Fig.~\ref{fig:qu.frac_v_m-star} shows the fraction of satellites at $z = 0$ that were quiescent prior to first infall.
The region boundaries are determined by our two parametrizations, with solid and dashed curves corresponding to those in Fig~\ref{fig:qu.frac_v_z}c.
If convolved with satellite infall times, the resultant quiescent-prior-to-infall fractions differ by less than 10\%, small compared with the marked difference from the quiescent fraction for central galaxies at $z = 0$ (red curve), which has been assumed for satellite initial conditions in previous works, as mentioned above.
Furthermore, the significant difference between satellite quiescent fractions at infall and at $z = 0$ (blue curve) indicates the importance of the satellite quenching process, which we will explore next.

\subsection{Importance \& efficiency of satellite quenching} \label{sec:quench_importance_efficiency}

We now explore the importance of satellite star formation quenching in building up the full population of quiescent (red-sequence) galaxies at $z = 0$ as well as the efficiency by which satellites are quenched.
The results in this subsection are essentially empirical, relying only on our parametrization for satellite initial quiescent fractions from \S\ref{sec:qu.frac-evol_cen}.
These results are independent of any particular model for the mechanism(s) or timescale of satellite quenching but provide insight into the efficiency of the mechanism(s) as a function of stellar mass.

The basic quantity that we use in this subsection is the number density of satellites (including ejected satellites) at $z = 0$ that quenched \tit{as satellites}, given by
\begin{eqnarray} \label{eq:quench_as_sat}
\nsatqassat(\mstar(z = 0)) & = & \nsatqnow(\mstar(z = 0)) - \\
& & \nsatqinf(\mstar(z = 0)) \, . \nonumber
\end{eqnarray}
Here, $\nsatqnow$ is the number density of satellites that are quiescent at $z = 0$ from the SDSS group catalog (again, using the simulation group catalog to correct for interloping galaxies), and $\nsatqinf$ is the number density of satellites at $z = 0$ that were quiescent prior to infall from our initial condition parametrization.
We express the latter quantity as a function of stellar mass at $z = 0$, based on our assumption that satellites have grown in stellar mass by the same amount as central galaxies (which we will justify in \S\ref{sec:m-star_growth}).

\subsubsection{Importance of satellite quenching} \label{sec:quench_importance}

\begin{figure}
\centering
\includegraphics[width = 0.99 \columnwidth]{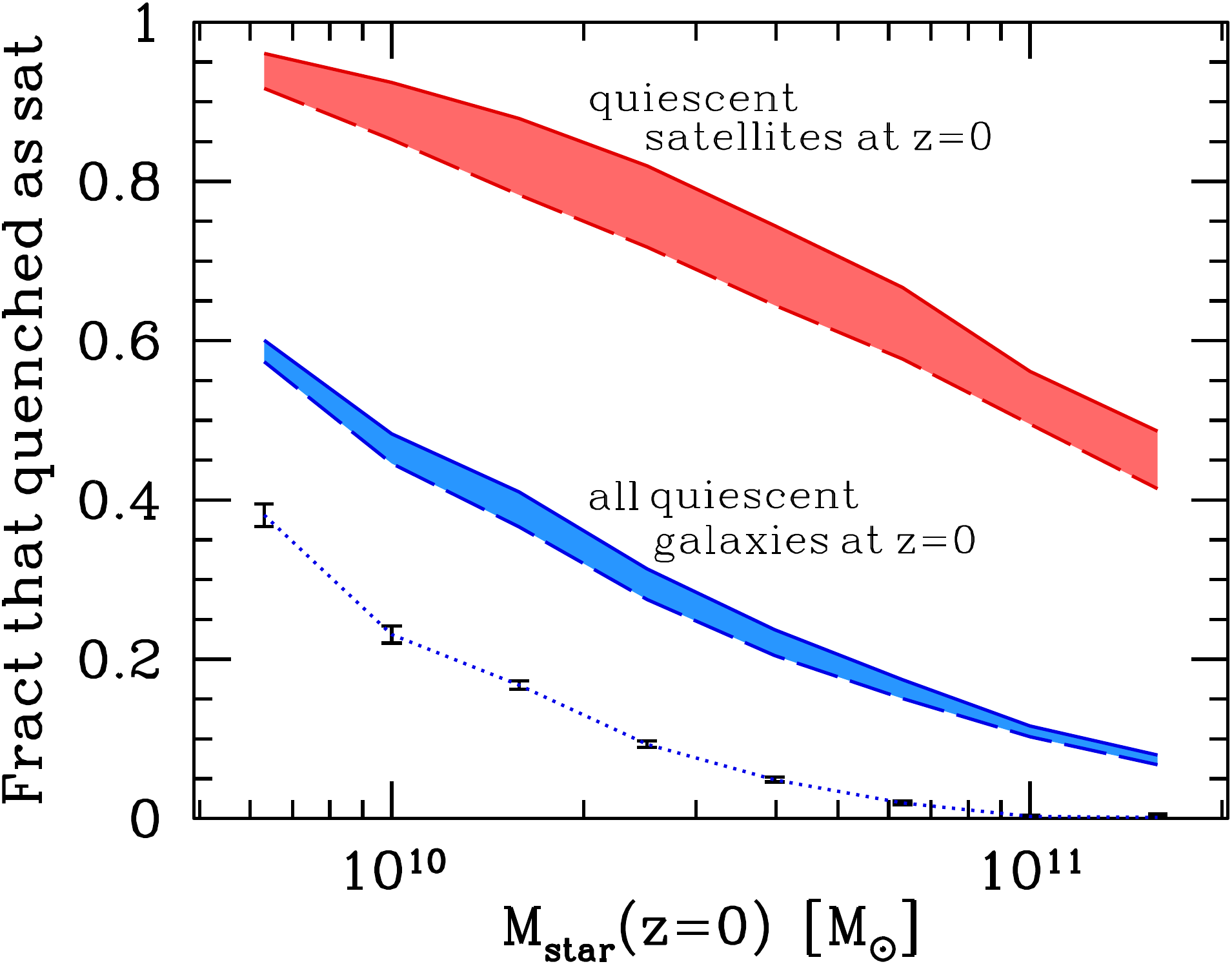}
\caption{
Importance of satellite quenching in building up the quiescent population at $z = 0$ as a function of stellar mass.
Top, red region shows the fraction of currently quiescent satellites that quenched as satellites, $\nsatqassat / \nsatqnow$.
Middle, blue region shows the fraction of \tit{all} currently quiescent galaxies that quenched as satellites, $\nsatqassat / \nallqnow$.
Region widths indicate uncertainty in satellite initial quiescent fractions from \S\ref{sec:qu.frac-evol_cen}.
Below $\mstar = 10 ^ {10} \msun$, satellite quenching is the dominant route for building up the entire quiescent (red-sequence) population.
Dotted blue curve shows the (incorrect) $\nsatqassat / \nallqnow$ from using central galaxies at $z = 0$ for satellite initial conditions.
} \label{fig:quench-as-sat.frac_v_m-star}
\end{figure}

We first examine the contribution of satellite quenching to building up the quiescent population at $z = 0$, as Fig.~\ref{fig:quench-as-sat.frac_v_m-star} shows.
First, to understand the importance of satellite quenching on just the satellite population, the red region shows what fraction of currently quiescent satellites were quenched as satellites, $\nsatqassat / \nsatqnow$.
The region width indicates the uncertainty from our two initial condition parametrizations in \S\ref{sec:qu.frac-evol_cen}.
At $\mstar < 10 ^ {10} \msun$, essentially all quiescent satellites quenched as satellites.
At $\mstar \sim 10 ^ {11} \msun$, this fraction is half, because half of satellites were already quiescent as central galaxies prior to infall (Fig.~\ref{fig:qu.frac_v_m-star}).
Thus, in our mass regime \tit{satellite quenching always dominates the production of quiescent satellites}.

To demonstrate the importance of satellite quenching on the entire galaxy population, the blue region shows what fraction of all currently quiescent galaxies quenched as satellites, $\nsatqassat / \nallqnow$.
This fraction decreases significantly with stellar mass, such that the vast majority of galaxy quenching occurs via central galaxies at $\mstar \gtrsim 10 ^ {11} \msun$.
In this regime, central galaxies are as likely as satellites to be quiescent at $z = 0$ (Fig.~\ref{fig:qu.frac_v_m-star}), and they outnumber satellites by a factor of $\gtrsim 6$ (Fig.~\ref{fig:qu.frac_v_z}b), so central galaxies dominate the production of the quiescent population.
By contrast, at the low mass end, even though central galaxies still outnumber satellites by a factor of $\sim 3$, satellite quenching is so much stronger that satellites dominate the production of the quiescent population.
Thus, \tit{at $\mstar < 10 ^ {10} \msun$, satellite quenching is the dominant route for the build-up of all quiescent (red-sequence) galaxies}.
Furthermore, no central (`isolated') galaxies are observed to be quiescent at $\mstar \lesssim 10 ^ {9} \msun$ \citep{GehBlaYan12}, so satellite quenching is the \tit{only} process for quenching galaxies at such low mass.

We emphasize the importance of accurate satellite initial conditions for these results.
Several previous works have attempted to infer the impact of satellite quenching by assuming that satellite initial conditions can be approximated via central galaxies at $z = 0$ \citep[e.g.,][]{vdBAquYan08, TinWet10, PenLilRen11, DeLWeiPog12}.
However, this assumption necessarily underestimates the importance of satellite quenching, because central galaxies were less likely to be quenched at higher redshift (Fig.~\ref{fig:qu.frac_v_z}c).
To highlight the impact of this assumption, the dotted blue curve in Fig.~\ref{fig:quench-as-sat.frac_v_m-star} shows the resultant $\nsatqassat / \nallqnow$ if one assumes (incorrectly) that central galaxy SFRs at $z = 0$ represent satellite initial conditions, that is, using $\nsatqinf = \fcenqnow \nsat$ in equation (\ref{eq:quench_as_sat}).
This assumption underestimates the true importance of satellite quenching by a factor of at least 50\% in our mass range.

\subsubsection{Efficiency of satellite quenching} \label{sec:quench_efficiency}

\begin{figure}
\centering
\includegraphics[width = 0.99 \columnwidth]{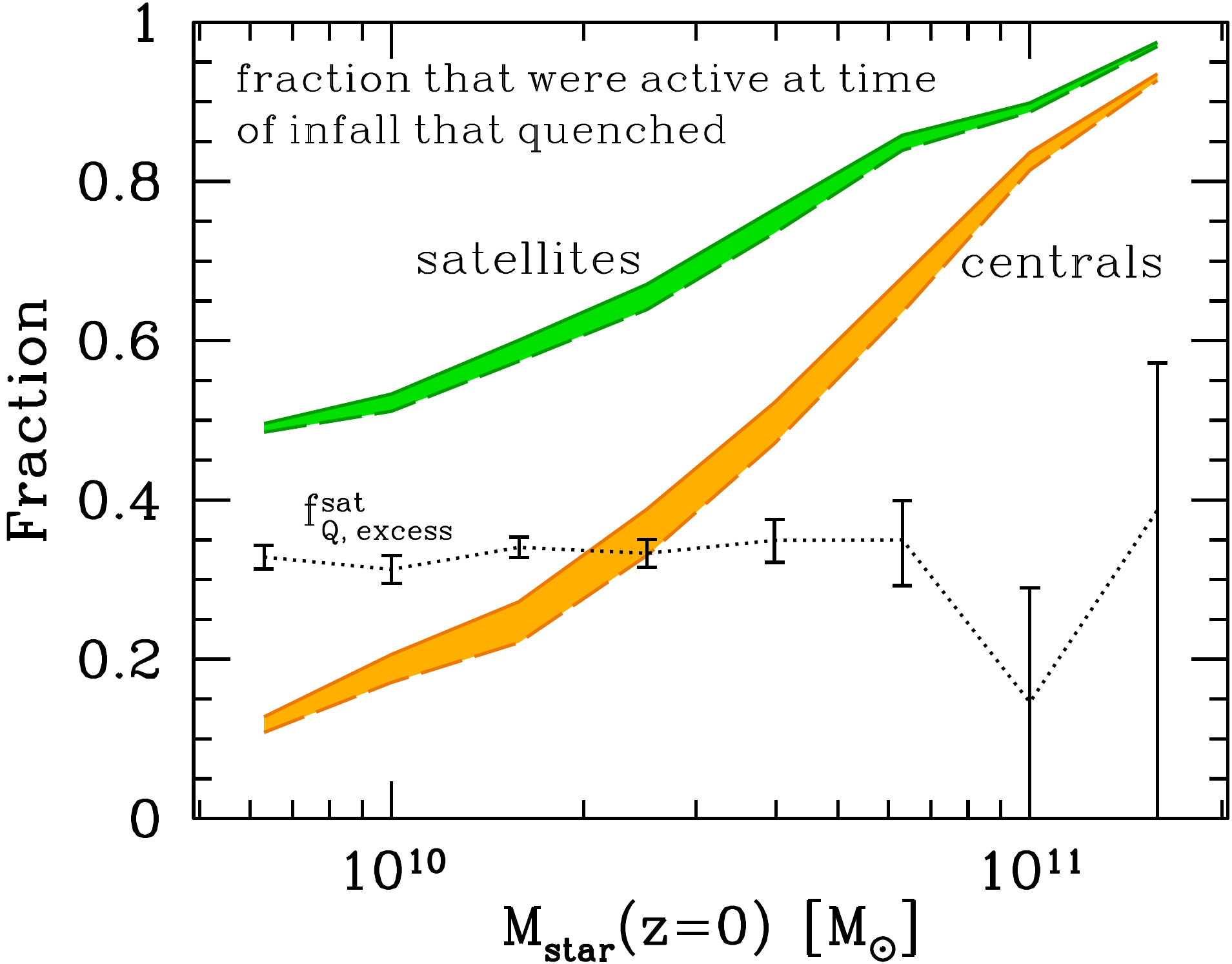}
\caption{
Satellite quenching efficiency versus stellar mass.
Top, green region shows the fraction of active-at-infall satellites that were quenched as satellites after infall, $\nsatqassat / \nsatainf$.
Region widths indicate uncertainty in satellite initial quiescent fractions from \S\ref{sec:qu.frac-evol_cen}.
More massive satellites are quenched more efficiently.
For reference, orange region shows fraction of active-at-infall satellites that would have quenched had they instead remained central galaxies.
Dotted curve shows the satellite quiescent fraction excess, $\fsatqexcess$, given by equation (\ref{eq:qu.frac_excess}).
} \label{fig:quench-after-infall-frac_v_m-star}
\end{figure}

We next examine the efficiency by which satellites are quenched, as given by the fraction of satellites that were active at infall (able to be quenched) that then quenched as satellites after infall, $\nsatqassat / \nsatainf$.
Fig.~\ref{fig:quench-after-infall-frac_v_m-star} (green region) shows this fraction as a function of stellar mass, with the width again indicating the uncertainty in satellite initial quiescent fractions.
At low mass ($\mstar < 10 ^ {10} \msun$) half of satellites that were active at the time of infall have been quenched by now, while the other half still actively form stars.
By contrast, at high mass ($\mstar > 10 ^ {11} \msun$) essentially all initially active satellites have been quenched.
Thus, \tit{more massive satellites are quenched more efficiently}.
Physically, this implies that more massive satellites are quenched more rapidly, as we will show in \S\ref{sec:quench_time_binary}.


To elucidate the differing quenching efficiencies for satellite versus central galaxies, the orange region in Fig.~\ref{fig:quench-after-infall-frac_v_m-star} shows what fraction of active-at-infall satellites would have quenched had they instead remained central galaxies.
This fraction is given by $\left( \fcenqnow - \fsatqinf \right) / \fcenanow$, with $\fsatqnow$ and $\fcenqnow$ being the fractions of satellite and central galaxies, respectively, that are quiescent at $z = 0$, and $\fcenanow = 1 - \fcenqnow$.
Like satellites, central galaxies also quench more efficiently at higher mass, though with a lower overall efficiency.
From Fig.~\ref{fig:quench-after-infall-frac_v_m-star}, it might naively appear that the differing quenching efficiency for satellite versus central galaxies is stronger at lower mass, though one must interpret these fractions carefully.
In \citetalias{WetTinCon12a}, we argued that a robust comparison is given by the satellite quiescent fraction excess
\begin{equation} \label{eq:qu.frac_excess}
\fsatqexcess = \frac{\fsatqnow - \fcenqnow}{\fcenanow} \,,
\end{equation}
which represents the excess fraction of satellites that were quenched after infall that would not have been quenched had they remained central galaxies.
As shown by the dotted curve in Fig.~\ref{fig:quench-after-infall-frac_v_m-star}, $\fsatqexcess$ is independent of stellar mass.
Furthermore, in \citetalias{WetTinCon12a} we showed that the stellar mass independence of $\fsatqexcess$ persists across all host halo masses and halo-centric radii.

If the same physical mechanism(s) that quenches central galaxies \tit{also} operates on satellites, then $\fsatqexcess$ indicates how much more of an effect the satellite-specific quenching mechanism(s) has.
In this scenario, the invariance of $\fsatqexcess$ suggests that the efficiency of the satellite quenching process(es) is independent of stellar mass.
However, the physical processes that are thought to quench central galaxies---such as virial shock heating, mergers, active galactic nuclei---and their relative importance as a function of stellar mass remain topics of active investigation.
Thus, it is unclear if such central galaxy quenching processes are important for satellites, and if they are, whether they occur before or after the onset of any satellite-specific processes.
We will investigate the physical mechanisms of satellite quenching in more detail in \citetalias{WetTinCon13b}, and we note that the results in this paper do not depend on the exact mechanism(s) at play.

To summarize this empirically-motivated subsection: (1) satellite quenching dominates the production of quiescent satellites at all masses we probe, and it dominates the production of \tit{all} quiescent galaxies at $\mstar < 10 ^ {10} \msun$, and (2) more massive satellites are quenched more efficiently.

\subsection{SFR evolution of satellites} \label{sec:sfr-evol_sat}

We now use the satellite infall times from our simulation to extend the results of the previous subsection and constrain satellite SFR evolution and quenching timescales.
This subsection presents the primary results of this paper.

The relative importance of various mechanisms for quenching satellites---such as strangulation, ram-pressure stripping, and harassment---and the details of how their effects propagate to influencing satellite star formation remain topics of active investigation.
Nonetheless, for most plausible physical processes, the likelihood that a satellite has been quenched increases with its time since infall.

Motivated by this idea, we proceed under the following ansatz: if a satellite was active at the time of first infall, \tit{the parameter that determines if it has been quenched is simply its time since first infall, $\tsinf$, the time that it has spent as a satellite.}
As we will show in \citetalias{WetTinCon13b}, this simple ansatz yields satellite quiescent fractions that have the correct dependencies on both halo-centric distance and halo mass, as compared with our observational results in \citetalias{WetTinCon12a}, because both the satellite quiescent fraction and $\tsinf$ increase with decreasing halo-centric distance.
This agreement implies that the scatter between quenching likelihood and $\tsinf$ must be small, because a scenario in which quenching likelihood and $\tsinf$ have large scatter would lead to satellite quiescent fraction radial gradients that are too shallow (see \citetalias{WetTinCon13b} for more).\footnote{
More rigorously, because we use a step-function threshold in $\tsinf$, selecting the maximally oldest surviving satellites to quench, any scatter in the relation between quenching and $\tsinf$ would lead to a characteristic quenching time (at which 50\% of active-at-infall satellites have quenched) that is necessarily shorter.
Thus, our quenching timescales are strictly upper limits, but, as we have argued, the scatter should be small.}
Thus, our ansatz provides a well-motivated means to constrain statistically the timescales over which satellites are quenched.

In this subsection, we examine satellite quenching timescales in two ways, the first being simpler and more empirical, the second being more physical.
First, in \S\ref{sec:quench_time_binary}, we consider quenching simply in the binary sense, that is, when SSFR falls below the $10 ^ {-11} \yrinv$ bimodality threshold.
We constrain the $\tsinf$ at which satellites are quenched in this binary sense, which we refer to as the satellite quenching time, $\tq$.
Then, in \S\ref{sec:quench_times}, we decompose this quenching time into two more physically informative timescales: the time delay after infall at which star formation \tit{starts} to be quenched, $\tqdelay$, and the characteristic e-folding time over which SFR fades once quenching has started, $\tauqfade$.

\subsubsection{Quenching timescale of satellites} \label{sec:quench_time_binary}

We first constrain the time since infall at which satellites are quenched (fall below $\ssfr = 10 ^ {-11} \yrinv$), $\tq$.
To do this, we select all satellites in the simulation at $z = 0$, and we use our satellite initial quiescent fraction parametrization from \S\ref{sec:qu.frac-evol_cen} to compute whether each was active or quiescent prior to infall.
Those that were quiescent remains so thereafter.
For those that were active at the time of infall, we designate the ones with $\tsinf > \tq$ as having been quenched.
Using narrow bins of both satellite and host halo mass, we adjust $\tq$ until the satellite quiescent fraction in the simulation group catalog matches that of the SDSS group catalog.
Repeating this procedure in each mass bin yields $\tq$ as a function of both satellite mass and host halo mass.

\begin{figure}
\centering
\includegraphics[width = 0.99 \columnwidth]{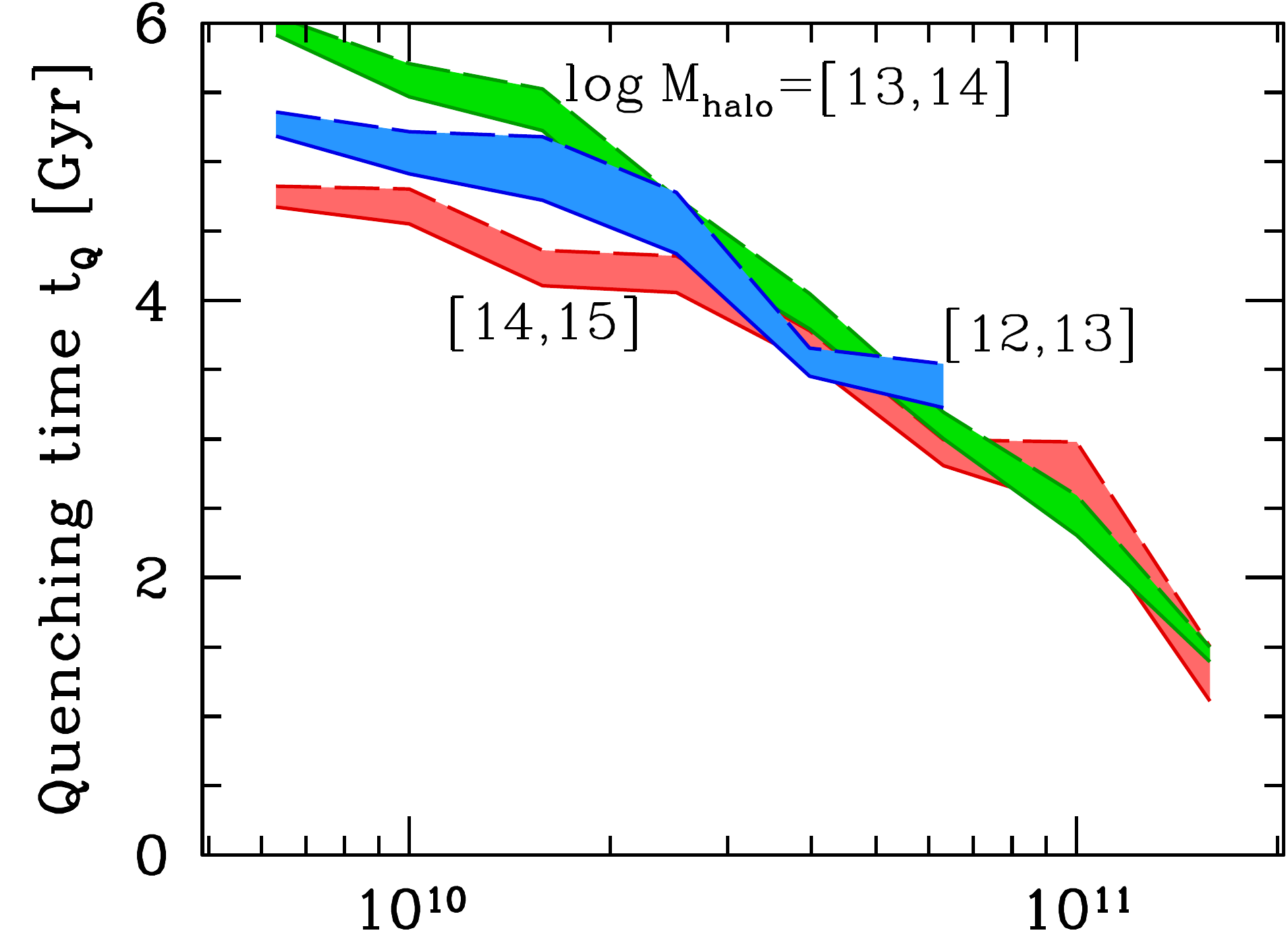}
\includegraphics[width = 0.99 \columnwidth]{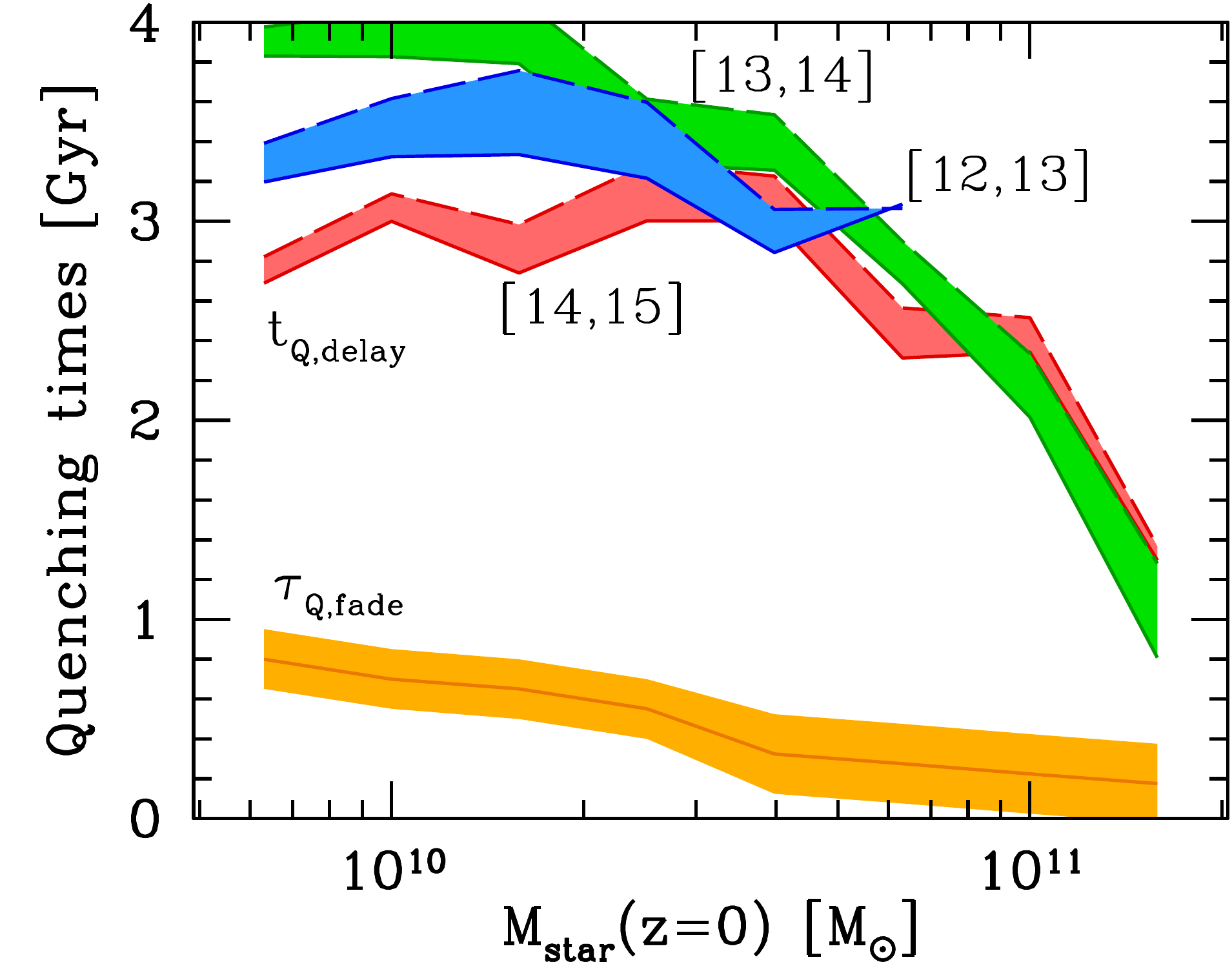}
\caption{
Satellite quenching timescales versus stellar mass.
\textbf{Top}: Time since first infall at which satellites are quenched (fall below $\ssfr = 10 ^ {-11} \yrinv$), $\tq$, in bins of their host halo's mass at $z = 0$.
Region widths indicate uncertainty in satellite initial quiescent fractions from \S\ref{sec:qu.frac-evol_cen}.
\textbf{Bottom}: Decomposing the above $\tq$ into the delay time after infall before satellite quenching starts, $\tqdelay$, and subsequent e-folding time over which SFR fades, $\tauqfade$.
Low-mass satellites take significantly longer to be quenched than those at higher mass, and their SFR fades more rapidly once quenching starts.
However, satellite quenching timescales do not depend on the mass of their host halo.
The uncertainty on $\tauqfade$ is dominated by uncertainty in fitting the full SSFR distribution.
} \label{fig:quench-time_v_m-star}
\end{figure}

Fig.~\ref{fig:quench-time_v_m-star} (top) shows how $\tq$ depends on satellites' current stellar mass, in bins of current host halo mass.
Region widths indicate the uncertainty from satellite initial quiescent fractions (solid and dashed curves correspond to those in Fig.~\ref{fig:qu.frac_v_z}c).
As the quenching efficiency results of \S\ref{sec:quench_efficiency} suggested, more massive satellites are quenched more rapidly.
The most massive satellites are quenched $\sim 2 \gyr$ after infall, while those at $\mstar < 10 ^ {10} \msun$ form stars actively for $\sim 5 \gyr$ after infall before being quenched.
Note that $5 \gyr$ is over half the age of the Universe at the typical redshift that they fell in, so low-mass satellites have spent as much as half of their entire star-forming lifetimes as satellites.

In \citetalias{WetTinCon12a}, we showed that satellites at $z = 0$ are more likely to be quiescent if they reside in more massive host halos.
But interestingly, in Fig.~\ref{fig:quench-time_v_m-star} we find no significant, systematic dependence of satellites' $\tq$ on their current host halo mass (there are fluctuations at low stellar mass, but they are not monotonic).
This halo mass independence naturally arises from our tying satellites' quenching to their time since \tit{first} infall, which incorporates the increased influence of group infall and ejection/re-infall in more massive halos, leading to a natural increase in $\tsinf$ with halo mass (Fig.~\ref{fig:t-inf_v_m-halo}).
Recall that this choice was motivated by the absence of a minimum host halo mass at $z = 0$ for affecting satellite SFR.
To the extent that this fact holds true out to $z \sim 0.5$ ($\sim 5 \gyr$ ago, the timescale over which $\tq$ is sensitive), Fig.~\ref{fig:quench-time_v_m-star} indicates that there is little-to-no freedom for $\tq$ to depend on host halo mass, because we obtain $\tq$ by empirically matching the observed quiescent fraction at $z = 0$ in each host halo mass bin.
In other words, given that a $\sim 10 ^ {12} \msun$ host halo quenches a low-mass satellite $\sim 4 \gyr$ after infall, as demanded by Fig.~\ref{fig:quench-time_v_m-star}, a similar halo that then falls into a massive cluster will bring in satellites that already are quenched.
Thus, the overall $\tq$ in massive clusters is set by a combination of $\tq$ from this group preprocessing and from satellites that fell directly into the cluster from the field, but as Fig.~\ref{fig:quench-time_v_m-star} shows, this overall $\tq$ is effectively the same for massive clusters as for a Milky-Way halo.\footnote{
Given that our SDSS group catalog constrains $\tq$ a function of host halo mass at $z = 0$ and not as a function of host halo mass at the time of infall directly, if group preprocessing were the primary mode of quenching satellites, this could mitigate the inferred dependence of $\tq$ on host halo mass.
However, as we will show in \S\ref{sec:where_quench}, most satellites quenched when they were in their current host halo, so any possible mitigation would be modest.}
Thus, \tit{the timescale over which satellites are quenched does not depend on the mass of their host halo}.

This lack of dependence on host halo mass may be surprising, given that more massive host halos represent more severe environments, having higher gas densities, temperatures, and satellite orbital velocities at a given scaled distance, $d / \rvir$.
But it is not clear that all possible satellite quenching processes should depend on host halo mass.
For example, if satellite quenching is driven simply by the inability to accrete gas after infall, then quenching occurs when a satellite exhausts its gas reservoir, independent of its host halo's mass.
Alternately, while the dominant satellite quenching process(es) may be more rapid at a given $d / \rvir$ in a more massive host halo, this is mitigated by the fact that dynamical friction causes a satellite of a given mass to orbit to smaller $d / \rvir$ more quickly in a lower mass host halo \citep{BoyMaQua08, JiaJinFal08}.
We will examine the dependence of satellite quenching on host halo mass with physically motivated models applied to orbital histories in \citetalias{WetTinCon13b}.

\subsubsection{`Delayed-then-rapid' quenching of satellites} \label{sec:quench_times}

While the satellite binary quenching timescale that we measured above, $\tq$, is advantageous in its simplicity, it is insensitive to the details of how satellite SFR evolves.
We now seek to understand satellite star formation histories more fully, as constrained by the full SSFR distribution at $z = 0$.

In \citetalias{WetTinCon12a}, we showed that the satellite SSFR distribution is bimodal, similar to central galaxies, across our stellar mass range.
The SSFR values of the active galaxy peak and bimodality break, as well as the fraction of galaxies near the bimodality break (`green valley'), do not vary with central versus satellite demarcation, host halo mass or halo-centric radius.
As we argued, these observations imply that (1) satellite SFRs evolve in the same manner as central galaxies for several Gyrs after infall, (2) the time since infall at which satellite SFR starts to be affected is long compared with the time over which SFR fades, and (3) the latter timescale does not depend on host halo mass or halo-centric radius.

To quantify these timescales, we build on these trends and construct a physically motivated, two-stage model for satellite SFR evolution.
The initial SFR for a satellite at its time of first infall, $\tinf$, is given by our parametrization in \S\ref{sec:sfr_at_infall}.
If a satellite was quiescent prior to infall, we do not evolve its SFR.
If a satellite was active at infall, we allow its SFR after infall to fade gradually in the same manner as central galaxies of the same stellar mass, using equation (\ref{eq:sfr-evol_cen}).
This central-type, gradual fading continues across a `delay' time, $\tqdelay$.
If $\tsinf > \tqdelay$, only then does a satellite start to be quenched, at which point we parametrize its SFR evolution via exponential fading, with $\tauqfade$ being the characteristic e-folding time over which SFR fades.
Defining $\tqstart = \tinf + \tqdelay$, satellite SFR evolves as 
\begin{equation} \label{eq:sfr-evol_sat}
\sfrsat(t) = 
\begin{cases}
\sfrcen(t) & t < \tqstart \\
\sfrcen(\tqstart) e ^ {\left\{ -\frac{(t - \tqstart)}{\tauqfade} \right\}} & t > \tqstart
\end{cases}
\end{equation}
to the redshift of our group catalog.
Note that $\tqdelay$ and $\tauqfade$ relate to $\tq$ from the previous subsection via $\tq = \tqdelay + N \tauqfade$, with $N = \ln \left[ \ssfr(\tqstart) / 10 ^ {-11} \yrinv \right]$.

Because the initial SFRs that we assign to satellites are based on observed distributions, any possible measurement uncertainty propagates into our resultant model SFRs at $z = 0$ as well, allowing for robust comparison with SDSS.
Also, for any satellite whose SSFR evolves below $\approx 10 ^ {-12} \yrinv$, we assign it as having $\ssfr = 10 ^ {-12} \yrinv$ plus log-normal scatter of 0.2 dex, which effectively mocks the measurement limits and scatter in the \citet{BriChaWhi04} method.

\begin{figure*}
\centering
\includegraphics[height = 0.25 \textheight]{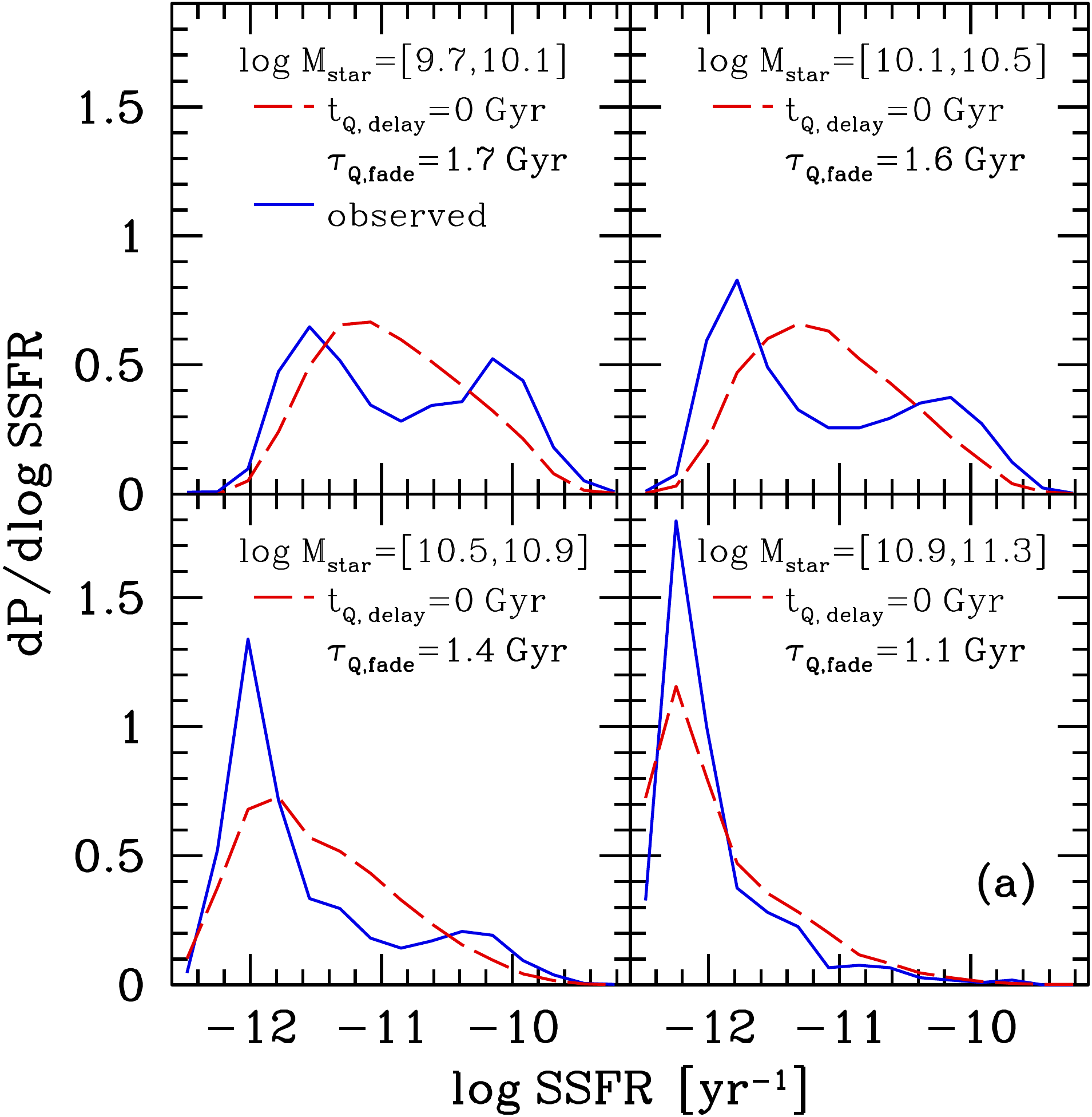}
\hspace{0.1cm}
\includegraphics[height = 0.25 \textheight]{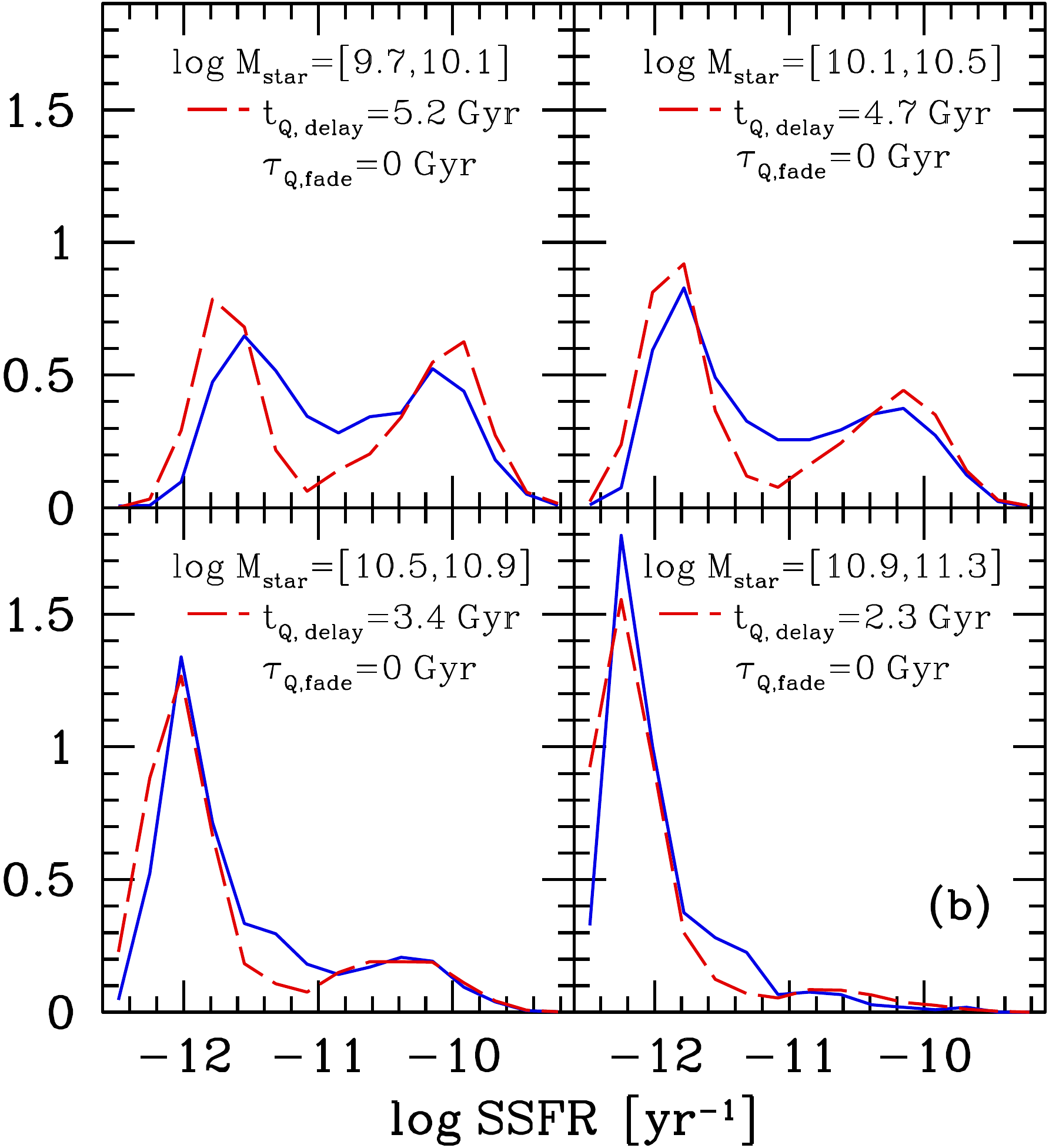}
\hspace{0.1cm}
\includegraphics[height = 0.25 \textheight]{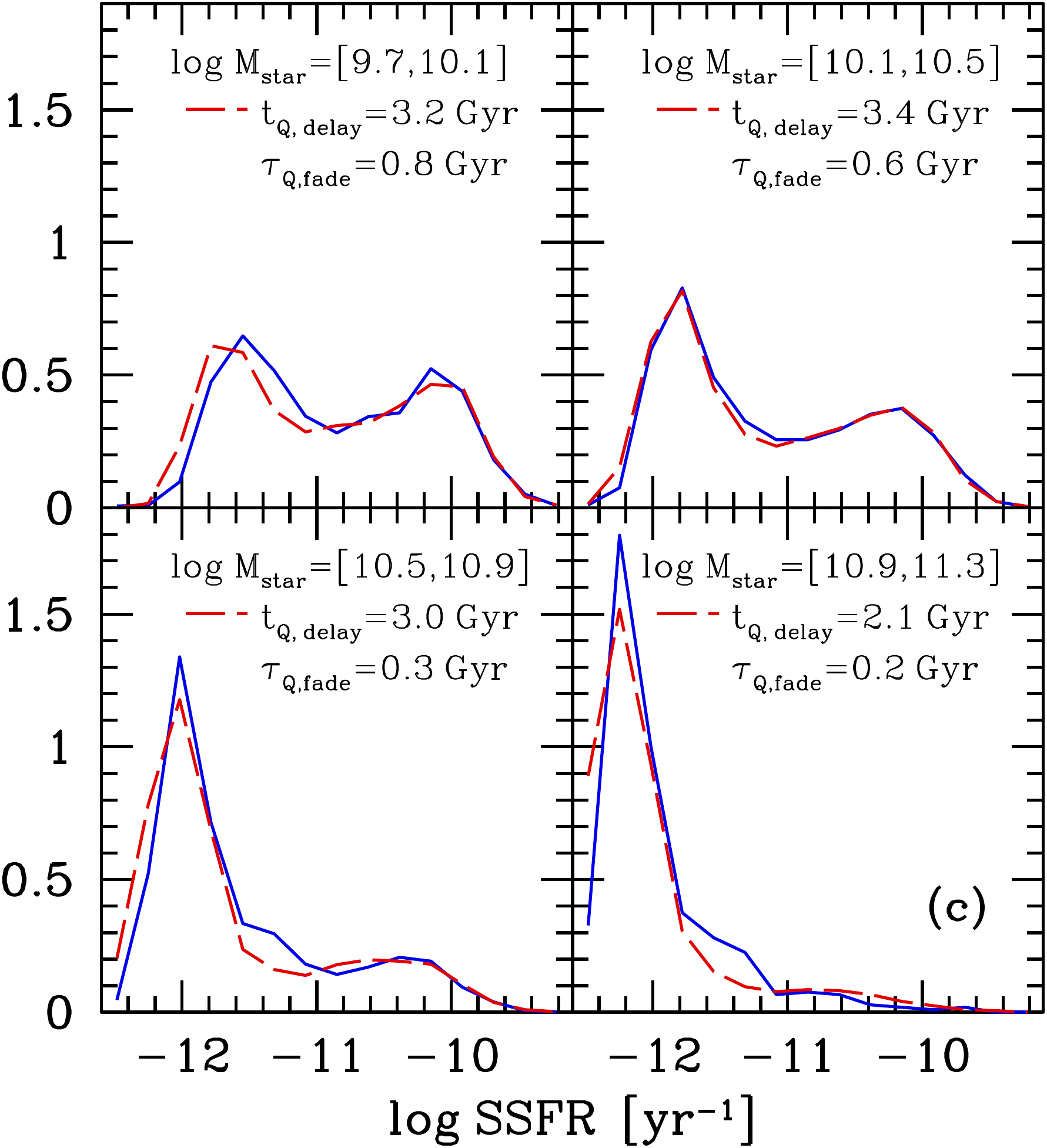}
\caption{
Specific star formation rate (SSFR) distribution of satellites at $z = 0$, in bins of stellar mass, in all host halos with $\mthm > 10 ^ {12} \msun$.
Solid blue curves show results from the SDSS group catalog while dashed red curves show results from the simulation group catalog using the satellite SFR evolution model of equation (\ref{eq:sfr-evol_sat}) under the following scenarios.
\textbf{(a)}: Satellite SFR starts to be quenched immediately upon infall and fades over an e-folding time, $\tauqfade$.
This scenario produces no bimodality.
\textbf{(b)}: Satellite SFR remains unaffected after infall for a quenching delay time, $\tqdelay$, after which it quenches instantaneously.
This scenario produces a bimodality that is too strong.
\textbf{(c)}: Satellite SFR remains unaffected after infall for time $\tqdelay$, after which it fades over an e-folding time $\tauqfade$.
This `delayed-then-rapid' quenching scenario produces the correct SSFR distribution.
These results are insensitive to host halo mass, and varying satellite initial quiescent fractions affects only $\tqdelay$ (see Fig.~\ref{fig:quench-time_v_m-star}).
} \label{fig:ssfr-distr_times}
\end{figure*}

Our physical, two-stage model for satellite SFR evolution has two free parameters (timescales) to constrain: $\tqdelay$ and $\tauqfade$.
We allow these timescales to vary, independently, with both satellite and host halo mass, again constrained to yield the observed satellite quiescent fraction at $z = 0$ in the joint mass bins.
The mere existence of a bimodal distribution with non-zero population at intermediate SSFRs requires that both timescales are non-zero.
However, to explore the impact of the two timescales, we consider three scenarios:
\begin{enumerate}
\renewcommand{\labelenumi}{(\alph{enumi})}
\item $\tqdelay = 0$, $\tauqfade$ fit to the quiescent fraction
\item $\tauqfade = 0$, $\tqdelay$ fit to the quiescent fraction
\item $\tqdelay$, $\tauqfade$ jointly fit to full SSFR distribution
\end{enumerate}
Fig.~\ref{fig:ssfr-distr_times} shows each resultant SSFR distribution at $z = 0$ in bins of stellar mass, for all satellites in host halos with $\mthm > 10 ^ {12} \msun$, using our fiducial parametrization for satellite initial quiescent fractions.
(Using our alternate parametrization shifts $\tqdelay$ to slightly longer values, as Fig.~\ref{fig:quench-time_v_m-star} (bottom) shows, but does not affect $\tauqfade$ or the quality of our fits.)
We discuss each scenario in turn.

First, we examine scenario (a), in which satellite quenching begins immediately at infall ($\tqdelay = 0$), and SFR fades slowly over a long $\tauqfade$.
Fig.~\ref{fig:ssfr-distr_times}a shows the resultant SSFR distributions that yield the correct satellite quiescent fractions.
This slow-fade scenario puts far too many satellites at intermediate SSFRs, violating the bimodality and leading to a qualitatively incorrect distribution.
\tit{Satellite SFR does not quench gradually after infall}.

Next, we consider the opposite scenario (b), in which satellite quenching is delayed after infall, but once it starts, it occurs instantly ($\tauqfade = 0$).
Fig.~\ref{fig:ssfr-distr_times}b shows the resultant SSFR distributions, which have a qualitatively correct bimodality, including a correct SSFR values of the active peak and bimodality break.
In particular, the agreement of the SSFR distribution for active galaxies confirms that they have evolved in the same manner as active central galaxies.
However, the bimodality break is clearly too strong, with a deficit of galaxies at intermediate SSFRs.
\tit{Satellite SFR must take non-trivial time to fade once quenching starts}.

Finally, in scenario (c) we allow both $\tqdelay$ and $\tauqfade$ to vary, as fit to the full SSFR distribution, providing a unique solution, as Fig.~\ref{fig:ssfr-distr_times}c shows.
This simple, two-stage quenching scenario produces a SSFR distribution in excellent agreement with observations at each stellar mass bin, with $\tauqfade$ being $10 - 20\%$ of $\tqdelay$.
In Fig.~\ref{fig:quench-time_v_m-star} (bottom), we show more explicitly how the best-fit $\tqdelay$ plus $\tauqfade$ times from Fig.~\ref{fig:ssfr-distr_times}c depend on current satellite mass, in bins of current host halo mass, including uncertainty in $\tqdelay$ from satellite initial quiescent fractions.
As with $\tq$ in the previous subsection, both $\tqdelay$ and $\tauqfade$ do not depend on the mass of the host halo.
We do not plot separate curves for $\tauqfade$ in bins of host halo mass, because the dominant uncertainty in $\tauqfade$, shown in Fig.~\ref{fig:quench-time_v_m-star}, comes from fitting the full SSFR distribution, which is larger than any systematic changes with host halo mass or satellite initial conditions.
Furthermore, as mentioned in \S\ref{sec:sfr-evol_cen}, we find that these $\tauqfade$ values do not depend on whether or not we evolve the SFR normalization for quiescent central galaxies.

To summarize this subsection, satellite SFR evolves, at least on average, via a `delayed-then-rapid' quenching scenario: satellite SFR remained unaffected for $2 - 4 \gyr$ after first infall (depending on stellar mass), after which SFR fades rapidly, with an e-folding time of $< 0.8 \gyr$.
Both timescales are shorter for more massive satellites have no significant dependence on host halo mass.

\section{Implications of satellite quenching timescales} \label{sec:implication}

Using the quenching timescales that we constrained in \S\ref{sec:quench_times}, we now explore two implications for satellite evolution.
First, in \S\ref{sec:where_quench} we explore where satellites were when they quenched, focusing on the importance of group preprocessing.
Second, in \S\ref{sec:m-star_growth} we use our constrained model for SFR evolution to examine how much satellites have grown in stellar mass since infall.

\subsection{Where were satellites when they quenched?} \label{sec:where_quench}

We have argued that the physical process(es) responsible for quenching satellites sets in at the time of first infall, but that its effects take considerable time to propagate before quenching starts.
We now ask: where were satellites at the moment when they started quenching?
In \S\ref{sec:infall_eject}, we examined what fraction of satellites fell in directly from the field versus as a satellite in a group.
We now extend those results, using the quenching times from \S\ref{sec:sfr-evol_sat}, to examine what fraction of currently quiescent satellite quenched (a) prior to first infall, (b) in a different host halo prior to falling into their current host halo, or (c) in their current host halo?

\begin{figure*}
\centering
\includegraphics[height = 0.198 \textheight]{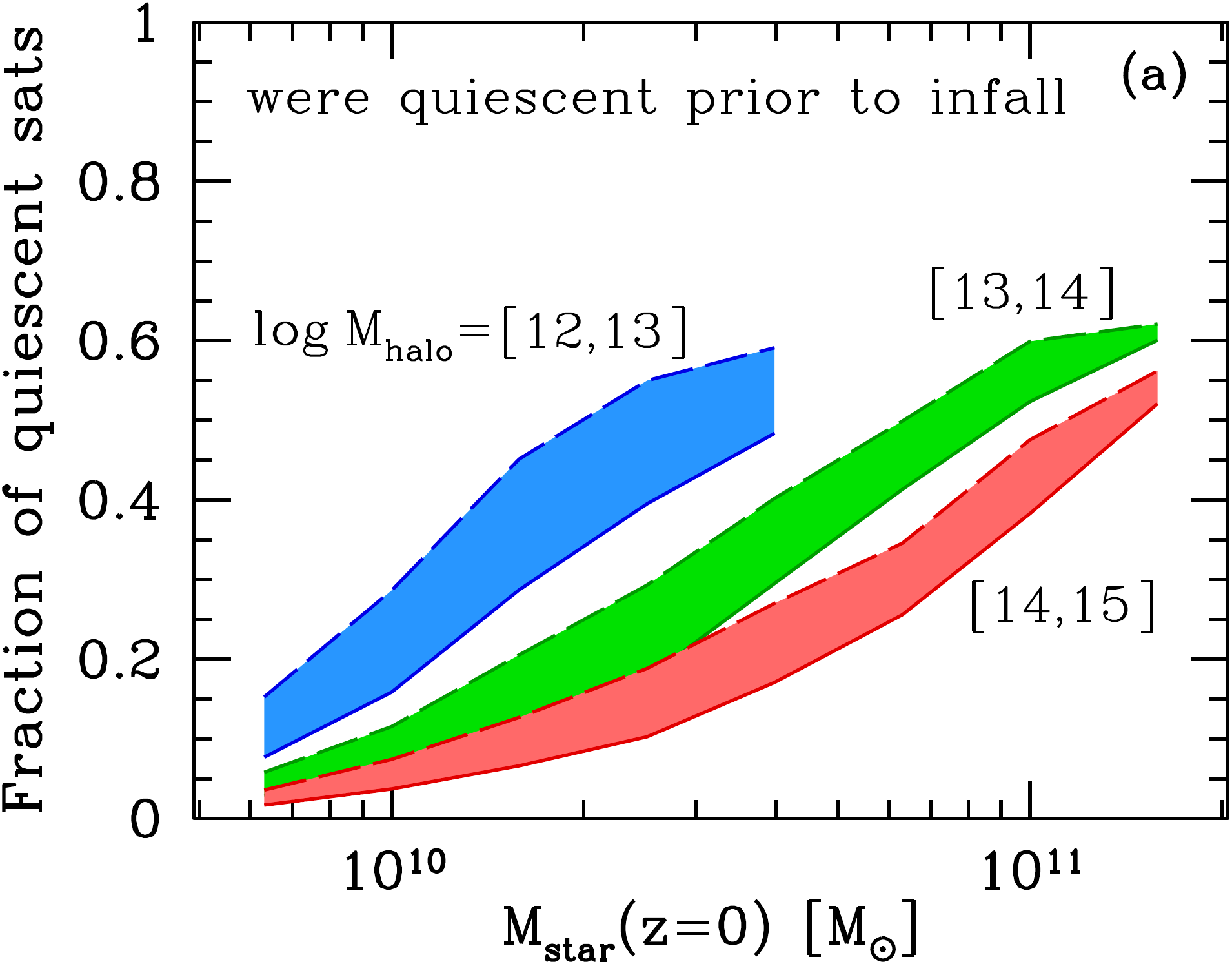}
\includegraphics[height = 0.198 \textheight]{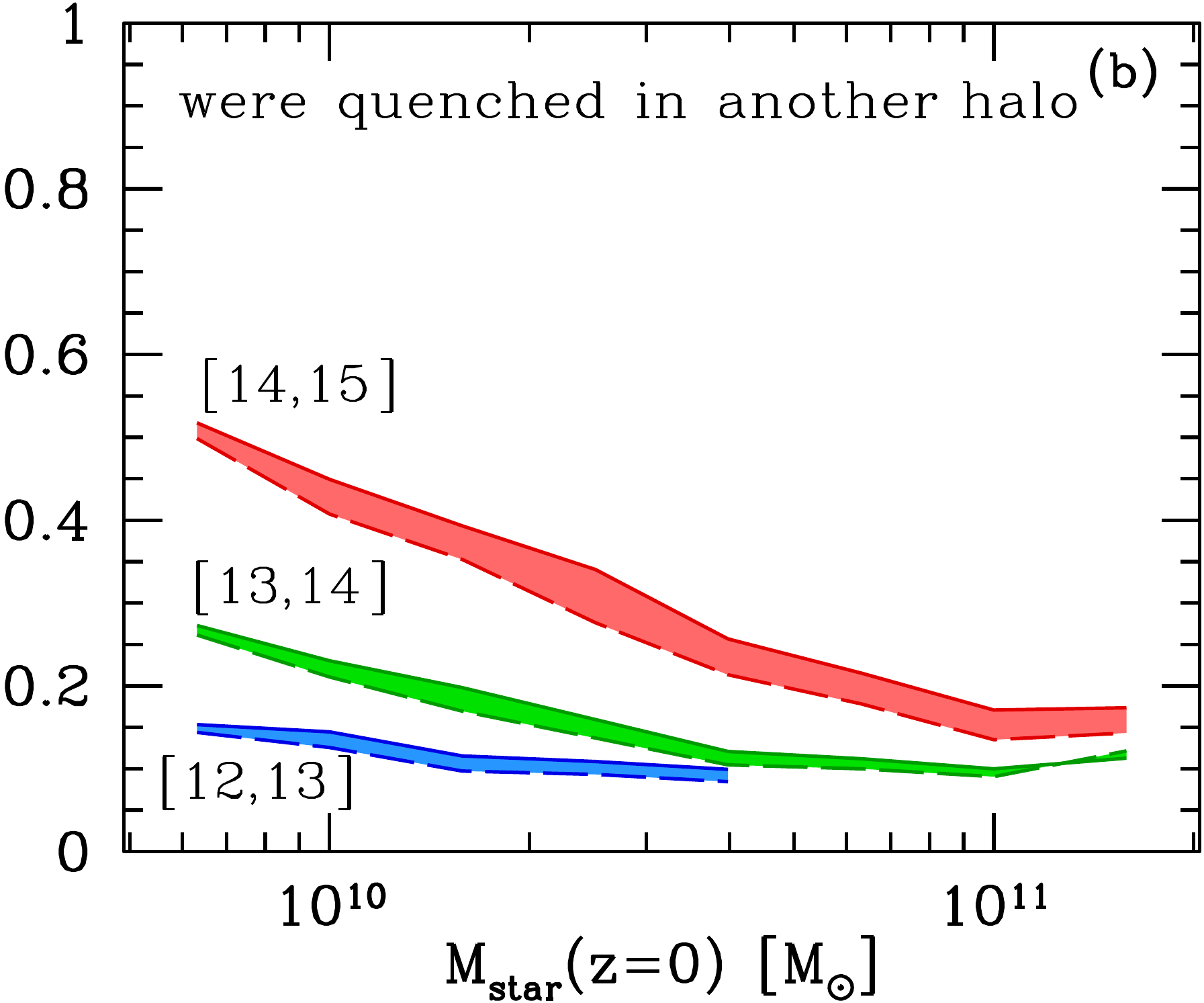}
\includegraphics[height = 0.198 \textheight]{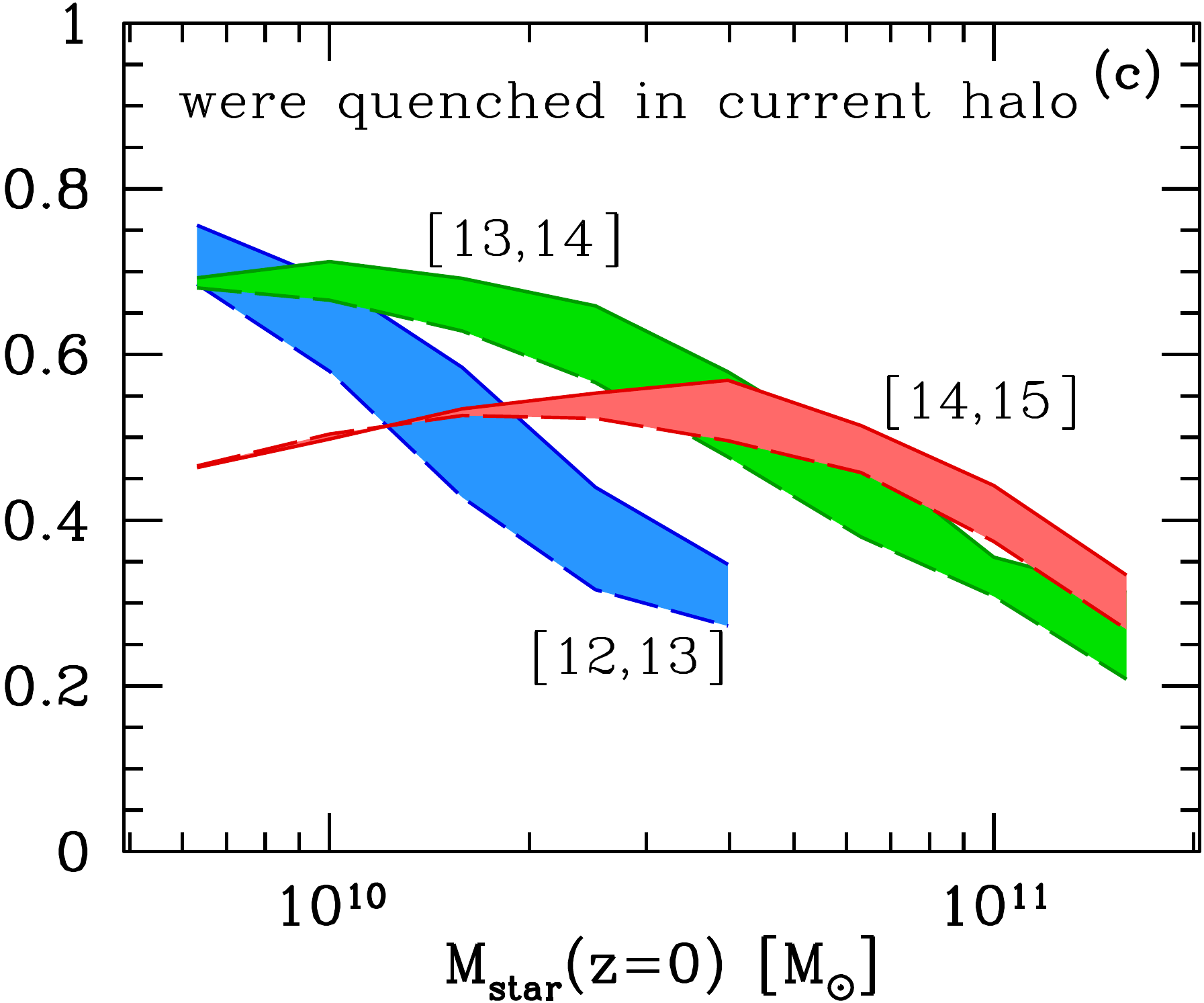}
\caption{
Where satellites were when they quenched: the fraction of currently quiescent satellites that quenched in different host halo regimes versus stellar mass, in bins of current host halo's mass.
Region widths indicate uncertainty in satellite initial quiescent fractions from \S\ref{sec:qu.frac-evol_cen}.
\textbf{(a)}: Fraction that were quenched as a central galaxy prior to first infall.
Quenching prior to infall is more important in regimes where satelites fell in more recently: higher mass satellites and in lower mass host halos (\S\ref{sec:infall-time_v_m}).
\textbf{(b)}: Fraction that started quenching in a different host halo (group) prior to falling into their current host halo, based on $\tqdelay$ from \S\ref{sec:quench_times}.
Half of quiescent satellites at $\mstar < 10 ^ {10} \msun$ that are currently in massive clusters began quenching as a satellite in a group, highlighting the importance of `group preprocessing'.
\textbf{(c)}: Fraction that quenched in their current host halo.
Overall, quenching prior to infall dominates at high satellite mass, group preprocessing is significant at low satellite mass, and quenching within the current host halo dominates at intermediate mass.
} \label{fig:where-quench_v_m-star}
\end{figure*}

For each surviving satellite in the simulation, we compute if it was active at the time of first infall, as before.
If so, we use the $\tqdelay$ values from \S\ref{sec:quench_times} for the time at which it started to quench.
We then compute whether the satellite was in the main progenitor of its current host halo or was in a different host halo at the time that quenching started.
Fig.~\ref{fig:where-quench_v_m-star} shows what fraction of currently quiescent satellites quenched in the three different regimes, as a function of current stellar mass, in bins of current host halo mass.

Fig.~\ref{fig:where-quench_v_m-star}a shows what fraction of currently quiescent satellites already were quenched as a central galaxy prior to first infall.
As discussed in \S\ref{sec:quench_importance_efficiency}, this fraction increases with stellar mass, both because the central galaxy quiescent fraction is higher at higher stellar mass and because higher mass satellites fell in more recently, when central galaxies were more likely to be quiescent.
Fig.~\ref{fig:where-quench_v_m-star}a also shows that quenching prior to first infall is more important in lower mass host halos, again because satellites in lower mass host halos fell in more recently.

Fig.~\ref{fig:where-quench_v_m-star}b shows what fraction of currently quiescent satellites started quenching in a different host halo prior to falling into their current host halo, indicating complete group preprocessing.
Opposite to the trends for quenching prior to first infall, this fraction is higher for lower mass satellites and in higher mass host halos, both trends a result of the hierarchical nature of halo growth (\S\ref{sec:infall-time}).
In particular, half of all low-mass ($\mstar < 10 ^ {10} \msun$) quiescent satellites in massive clusters ($\mthm > 10 ^ {14} \msun$) started quenching as a satellite in a group.

Finally, Fig.~\ref{fig:where-quench_v_m-star}c shows what fraction of currently quiescent satellites quenched while in their current host halo.
This mode of quenching dominates at most masses, though the fraction is alway $\lesssim 70\%$; its importance wanes both at high stellar mass, where quenching prior to first infall dominates, and at low stellar mass, where the importance of group preprocessing increases.
Only about half of quiescent satellites within massive clusters ($> 10 ^ {14} \msun$) quenched there.

In summary, group preprocessing has a critical impact on satellite star formation histories.
We have argued that \tit{any} time spent as a satellite in another host halo is important as far as starting the quenching process, but these results demonstrate the impact of complete group preprocessing.
This is particularly important for satellites in clusters, in which $15 - 50\%$ of all quiescent satellites started quenching as a satellite in another host halo.
Given the hierarchical nature of halo growth, group preprocessing should be only more important for quenching satellites below our $5 \times 10^9 \msun$ stellar mass limit.


\subsection{Stellar mass growth after infall} \label{sec:m-star_growth}

So far, we examined satellite star formation histories with a focus on star formation \tit{rate} evolution, but in this last subsection we examine the implications for stellar mass growth.
In \S\ref{sec:sfr-evol_sat}, we showed that satellites continue to form stars actively, in the same manner as central galaxies, for $2 - 4 \gyr$ after infall, which represents as much as half of their total star-forming lifetimes.
Thus, satellites have the capacity to grow significantly in stellar mass via star formation after infall.
For now, we ignore any other processes that might affect satellite stellar mass evolution, such as tidal stripping or merging, though we discuss these in Appendix \ref{sec:mass_growth_sham}.

\begin{figure}
\centering
\includegraphics[width = 0.99 \columnwidth]{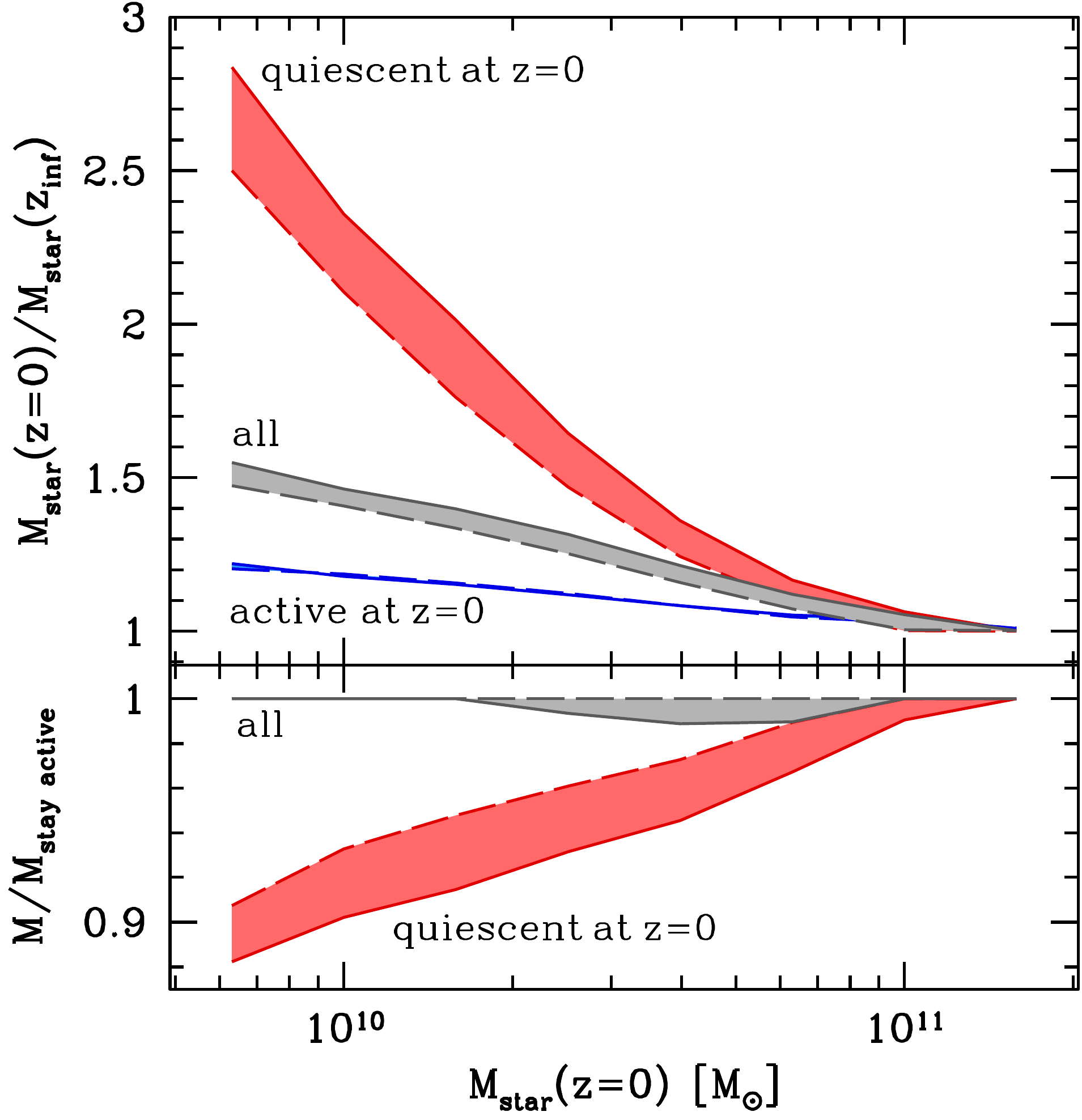}
\caption{
Satellite stellar mass growth after infall.
\textbf{Top}: Median ratio of a satellite's stellar mass at $z = 0$ to what it had at first infall versus current stellar mass, for all satellites (middle, grey), those that remain active (bottom, blue) and those that are quiescent (top, red) at $z = 0$.
Region widths show uncertainty in satellite initial quiescent fractions from \S\ref{sec:qu.frac-evol_cen}.
\textbf{Bottom}: Median ratio of a satellite's stellar mass at $z = 0$ to what it would have at $z = 0$ had it not quenched after infall.
Satellite quenching has little impact on stellar mass growth: satellite and central galaxy stellar mass growths via star formation are nearly identical.
} \label{fig:m-star-growth_v_m-star}
\end{figure}

To quantify the amount of stellar mass growth via star formation that satellites at $z = 0$ have experienced since first infall, we use our model for satellite SFR evolution given by equation (\ref{eq:sfr-evol_sat}), with the appropriate values of $\taucen$ and the quenching timescales $\tqdelay$ and $\tauqfade$ from \S\ref{sec:quench_times} given a satellite's stellar mass at $z = 0$.
We integrate $\sfr(t)$ to obtain the stellar mass formed since first infall, again assuming that 40\% of this stellar mass is lost through supernovae and stellar winds.
For a satellite that was quiescent prior infall, its SFR has been sufficiently low that we can neglect any stellar mass growth since that time.
We then examine statistical trends by computing the median fractional stellar mass growth since first infall in bins of stellar mass at $z = 0$.

Fig.~\ref{fig:m-star-growth_v_m-star} (top) shows the median ratio of a satellite's stellar mass at $z = 0$ to the mass that it had at the time of its first infall, as a function of its current stellar mass.
As before, region widths show uncertainty in satellite initial quiescent fractions from \S\ref{sec:qu.frac-evol_cen}.
Considering all surviving satellites (grey region), their median stellar mass growth since infall is negligible at high mass but is 50\% at $\mstar < 10 ^ {10} \msun$.
This mass dependence arises because lower mass satellites are more likely both to have fallen in earlier when SFRs were higher and to have been active at the time of infall.

The blue and red regions in Fig.~\ref{fig:m-star-growth_v_m-star} show median values for currently active and quiescent satellites, respectively.
Overall, currently active satellites have experienced significantly less stellar mass growth since infall than currently quiescent galaxies.
While perhaps counter-intuitive, this trend is readily understandable.
Even though active satellites are still growing in stellar mass, to remain active they necessarily fell in more recently, meaning both lower SFRs at the time of infall and less time for mass growth after infall.
To understand currently quiescent satellites, note that they are composed of two populations: those that were quiescent prior to infall and those that quenched after infall.
While those that were quiescent prior to infall did not grow in stellar mass at all, those that quenched after infall necessarily fell in early, when SFRs were much higher, and they then spent several Gyrs actively forming stars before being quenched.
While the former population dominates at high mass, the latter dominates at low mass, leading to stellar mass typically having more than doubled since infall at $\mstar < 10 ^ {10} \msun$.
Thus, \tit{low-mass satellites experience considerable stellar mass growth after infall.}

To put this result in context, we compare the stellar mass growth experienced by satellite versus central galaxies.
Fig.~\ref{fig:m-star-growth_v_m-star} (bottom) shows the median ratio of satellite mass at $z = 0$ to what it would be if all satellites that were active at infall remained active to $z = 0$, that is, if satellites never quench.
This approximately indicates the ratio of stellar mass that satellites have compared to what they would have if they had remained a central galaxy.
(In fact, the masses of satellite versus central galaxies are closer than in Fig.~\ref{fig:m-star-growth_v_m-star}, because some central galaxies have quenched since the time that a satellite fell in, but we do not attempt to fully model central galaxy SFR evolution here.)
While this mass ratio is unity for currently active satellites (by definition), considering the entire satellite population, the median ratio remains consistent with unity (grey region).
Considering just currently quiescent satellites, which have experienced the most truncated mass growth, the median reduction in stellar mass at $z = 0$ is never more than 10\% (red region).
Furthermore, considering just those satellites that quenched \tit{after} infall, the reduction is still only 10\%, though it remains at that level across all stellar mass (not shown).
Thus, despite the clear importance that satellite quenching has on instantaneous SFR, \tit{satellite-specific quenching has minimal impact ($\lesssim 10\%$) on satellite stellar mass growth; satellite and central galaxy stellar mass growth via star formation is nearly identical}.
This behavior arises for two reasons: satellites evolve for considerable time ($2 - 4 \gyr$) after infall until they start to be quenched, and galaxies in our mass range form the vast majority of their stars at high redshift ($z \gtrsim 0.5$) when their SFRs were much higher, so quenching at low redshift has little impact on their final stellar mass.

As outlined in \S\ref{sec:sfr_at_infall}, in order to assign accurate initial SFRs to satellites at their time of first infall, we estimated their stellar mass at that time via the ansatz that they grew by the same amount as central galaxies of the same stellar mass.
This is not obviously a good approximation, for instance, if observations had constrained both $\tqdelay$ and $\tauqfade$ in \S\ref{sec:quench_times} to be quite short.
However, the results of this subsection reassuringly show that our approach is statistically self-consistent to good approximation.
Moreover, we investigated alternative scenarios in which satellites have grown significantly less in stellar mass since infall than central galaxies, but this generically leads to even longer quenching times, and thus more implied stellar mass growth, so these scenarios are not internally self-consistent.
Thus, based on Fig.~\ref{fig:m-star-growth_v_m-star}, the only self-consistent scenario is that stellar mass growth in satellites is the same as central galaxies to within 10\% (at least in the absence of significant, systematic stellar mass loss from tidal stripping).

Finally, the results of this section provide physical insight and possible improvement into the implementation of the SHAM method in assigning galaxy stellar mass to \tit{both} central and satellites subhalos.
The basic idea of SHAM is that one can assign instantaneous stellar mass to all subhalos, central and satellite, based simply on some measure of their subhalo mass (or circular velocity).
Our results show that, because satellites and central galaxies grow in stellar mass by essentially the same amount, on average, it is justifiable to assign stellar mass to all subhalos under a single, simple prescription.
However, our results do imply possible tension with the \tit{way} that SHAM typically is implemented, specifically, through the use of the maximum/infall subhalo mass, which does not evolve after infall for satellites.
We discuss this issue in Appendix \ref{sec:mass_growth_sham}.

\section{Summary and Discussion} \label{sec:summary}

\subsection{Summary}

\begin{figure}
\centering
\includegraphics[width = 0.99 \columnwidth]{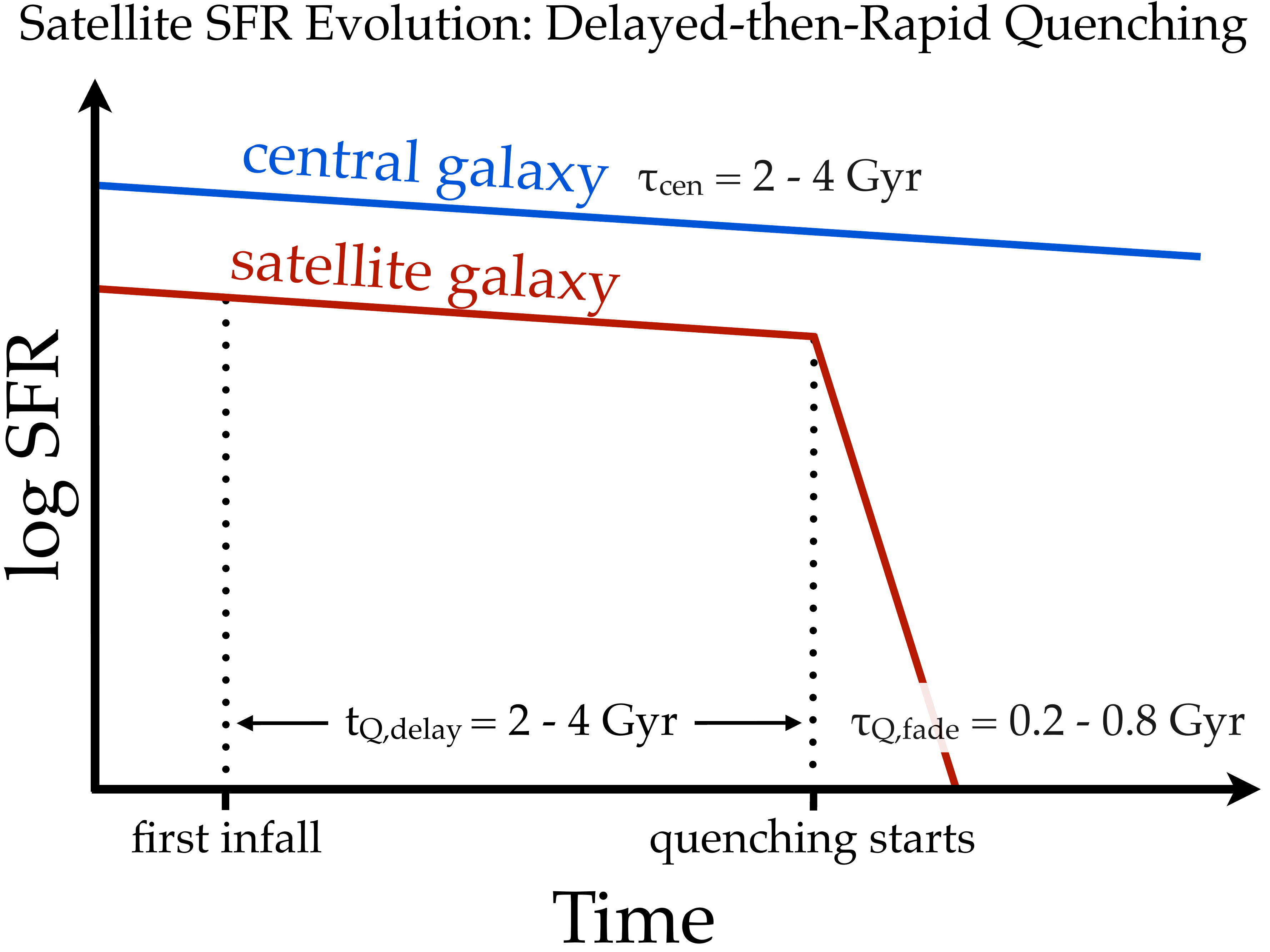}
\caption{
Summary diagram of satellite galaxy SFR evolution, as given by equation (\ref{eq:sfr-evol_sat}), highlighting the `delayed-then-rapid' quenching scenario.
The physical process(es) responsible for quenching star formation in a satellite begins after it first falls into another host halo, regardless of the host halo's mass.
However, it takes considerable time for SFR to be affected: satellite SFR evolves after infall in the same manner as central galaxies for a delay time $\tqdelay = 2 - 4 \gyr$ (red curve), depending on stellar mass.
Then, the satellite's star formation starts to be quenched, and SFR fades rapidly, with an e-folding time $\tauqfade = 0.2 - 0.8 \gyr$, also depending on stellar mass.
Less massive satellites have longer $\tqdelay$ and $\tauqfade$, but neither timescale dependends on the mass of the host halo.
Because of the long $\tqdelay$, satellite stellar mass growth via star formation is nearly equal to that of central galaxies.
Because of hierarchical halo growth, many satellites in massive host halos were quenched as a satellite in a lower mass halo prior to infall.
For comparison, blue curve shows gradual SFR fading for central galaxies, with characteristic fading time $\taucen = 2 - 4 \gyr$ from equation (\ref{eq:sfr-evol_cen}).
} \label{fig:sfr-evol_diagram}
\end{figure}

Using a galaxy group/cluster catalog from SDSS Data Release 7, together with a cosmological \tit{N}-body simulation to track satellite orbits, we examined in detail the star formation histories of satellite galaxies at $z \approx 0$, focusing on their times since infall, quenching timescales, and stellar mass growth after infall.
Applying the same group-finding algorithm to our simulation as we used in SDSS allows us to make robust comparisons of model results to observations.
To obtain accurate initial conditions for the SFRs of satellites at their time of first infall, we constructed an empirically based, statistical parametrization for the evolution of central galaxy SFRs out to $z = 1$; this is critical for the accuracy of our results because, at fixed stellar mass, the quiescent fraction for central galaxies more than doubles from $z = 1$ to 0.
Our primary result is that, at least on average, satellite SFR evolves via a `delayed-then-rapid' quenching scenario: satellite SFR remained unaffected for several Gyrs after first infall, after which quenching occurs rapidly, as Fig.~\ref{fig:sfr-evol_diagram} summarizes.
In more detail, our main results are:

\tit{Infall as part of a group and ejection beyond $\rvir$ are important aspects of satellite evolution.}
Fewer than half of satellites in massive clusters fell in directly from the field; the rest fell in as a satellite in another host halo or experienced secondary infall after becoming ejected.
Satellites at $z = 0$ experienced their first infall typically at $z \sim 0.5$, or $\sim 5 \gyr$ ago, with a broad tail out to $z \ge 1$.
Less massive satellites and those in more massive host halos fell in earlier.

\tit{Satellite quenching is a critical process of galaxy evolution.}
Satellite quenching always dominates the production of quiescent satellites.
Moreover, satellite quenching is responsible for producing the majority of \tit{all} quiescent (red-sequence) galaxies at $\mstar < 10 ^ {10} \msun$ by $z = 0$.

\tit{Satellite quenching is delayed-then-rapid.}
Based on observations, we argued that the process(es) responsible for quenching satellites begins after \tit{first} infall into any other host halo.
As constrained by the satellite SSFR distribution at $z = 0$, satellites, at least on average, then remain actively star-forming for $2 - 4 \gyr$ (depending on stellar mass) after first infall, unaffected by their host halo, before quenching starts.
Once quenching has started, the e-folding time over which SFR fades is $< 0.8 \gyr$.

\tit{Satellite quenching timescales are shorter at higher stellar mass but are independent of host halo mass.}
More massive satellites start to be quenched more rapidly after infall, and once quenching has started, their SFRs fade more quickly.
However, these quenching timescales do not depend on host halo mass.
The observed increase in the satellite quiescent fraction with host halo mass arises because of the increased importance of group preprocessing and ejection/re-infall in more massive host halos.

\tit{Group preprocessing plays a critical role in quenching satellites.}
Half of low-mass ($\mstar < 10 ^ {10} \msun$) quiescent satellites in clusters started quenching in another host halo before falling into the cluster.
Across all satellite masses, the fraction of quiescent satellites that were quenched  within their current host halo is never more than $\sim 70\%$.

\tit{Satellite quenching barely impacts stellar mass growth.}
Because satellite quenching is so delayed, low-mass satellites have experienced considerable stellar mass growth via star formation since infall: satellites at $\mstar < 10 ^ {10} \msun$ are, on average, 50\% more massive than at infall.
Moreover, the average amount of mass growth via star formation in satellite and central galaxies is identical to within 10\%.
This provides key physical insight into the abundance matching technique for assigning stellar mass to subhalos, as outlined in Appendix \ref{sec:mass_growth_sham}.

\subsection{Relation to satellite gas content} \label{sec:quench_time_gas}

We first discuss the relation of our results to satellite gas content.
Satellites provide unique laboratories for examining gas depletion and its relation to star formation because, unlike central galaxies, satellites' subhalos are thought not to accrete matter after infall: the strong gravitational tidal forces in the host halo both prevent a satellite's subhalo from accreting new matter and strip any existing subhalo matter, including gas, from the outside-in.
Additionally, any thermalized gas in the host halo can heat and ram-pressure strip any extended subhalo gas, and in the extreme case of both high gas density and satellite velocity, ram-pressure can strip cold gas directly from the disc.

We first discuss the implications of our $\tqdelay$ results, that is, that SFR in satellites evolves for $2 - 4 \gyr$ after infall, depending on stellar mass, \tit{unaffected by the host halo}.
While this timescale may represent, to some degree, the statistical average of those of individual satellites, it is informative to consider in the context of gas depletion times.
Given that star formation is fueled by cold, molecular gas \citep{WonBli02, BigLerWal08}, this means that a significant quantity of such gas must persist in a satellite's disc for that amount of time.

One possibility is that a sufficient reservoir of cold gas was present in the disc at the time of infall.
As a constraint, we compare our $\tqdelay$ times to observed cold gas depletion times, defined as $M_{\rm gas} / \sfr$.
At $z = 0$, observed atomic gas depletion times in $\mstar > 10 ^ {10} \msun$ galaxies are $\sim 3 \gyr$, with large scatter but no systematic dependence on stellar mass or SFR \citep{SchCatKau10}.
Incorporating the additional $\sim 30\%$ of the gas that is molecular \citep{SaiKauKra11}, the total gas depletion time would extend to $\sim 4 \gyr$.
This timescale can be even longer to the extent that gas recycled from stellar mass loss fuels star formation \citep[e.g.,][]{LeiKra11}.
If valid at higher redshift, this depletion time would be sufficient to accommodate our $\tqdelay$ values \tit{if} all of the atomic gas converts to stars.

Observed gas ratios, $M_{\rm gas} / \mstar$, provide another constraint.
In \S\ref{sec:m-star_growth}, we showed that satellites experience significant stellar mass growth via star formation.
In particular, currently quiescent, low-mass satellites have more than doubled their stellar mass since infall, which requires a gas reservoir comparable in mass to their stars at the time of infall.
Observations at $z = 0$ show that actively star-forming galaxies at $\mstar \sim 10 ^ {10} \msun$ have total cold gas masses that are $\sim 40\%$ of their stellar mass \citep{CatSchKau10, SaiKauKra11}, and that this gas ratio increases with decreasing stellar mass, being near unity for galaxies just below our mass threshold \citep[e.g.,][]{GehBlaMas06}.
However, currently quiescent, low-mass satellites fell in at higher redshift (typically, $z = 0.5 - 1$), and if the total cold gas fraction increases with redshift at a rate suggested by observations of molecular gas in actively star-forming, massive ($\mstar > 10 ^ {10} \msun$) galaxies at $z = 0.4 - 1.4$ \citep{DadBouWal10, TacGenNer10, GeaSmaMor11}, then these satellites would have had enough cold gas in their disc to accommodate the significant stellar mass growth in Fig.~\ref{fig:m-star-growth_v_m-star}.

Furthermore, the cold gas in the disc could be replenished for some time after infall if the most concentrated and tightly-bound component of the extended subhalo gas continues to cool/accrete onto the disc for several Gyrs before being stripped.
X-ray observations show that roughly half of massive satellites in groups and clusters retain extended, hot gas halos, though truncated as compared with central (`field') galaxies \citep{SunDonVoi07, JelBinMul08}.
Simulations also show retention of extended subhalo gas:
\citet{McCFreFon08} found that satellite subhalos can retain a significant fraction ($\sim 30\%$) of their hot gas for several Gyrs after infall, while \citet{SimWeiDav09} and \citet{KerKatFar09} found that satellites continue to accrete significant gas onto their disc, though at a reduced rate compared with central galaxies.
Adding this replenishment to what cold gas was already in the disc at infall, the total gas reservoir in/around satellites appears fully sufficient to fuel their extended SFR and stellar mass growth as demanded by $\tqdelay$.

Finally, regarding $\tqdelay$, we note that the halo radius crossing time, given our virial definition, is $t_{\rm cross} = \rvir / \vvir = 2.7 (1 + z) ^ {-3 / 2} \gyr$, independent of host halo mass.
More precisely, numerically integrating satellite orbits in an NFW potential (assuming energy and angular momentum conservation) using typical initial orbital parameters from \citet{Wet11} for satellites in host halos in our mass range, the average time from infall to first pericentric passage is somewhat shorter at $2 (1 + z) ^ {-3 / 2} \gyr$, independent of satellite mass.
Thus, the onset of satellite quenching occurs near the time of pericentric passage for more massive satellites and $1 - 2 \gyr$ after pericentric passage for lower mass satellites, as we will explore in more detail in \citetalias{WetTinCon13b}.

We next discuss the implications of our $\tauqfade$ results, that is, that once quenching has started, satellite SFR fades rapidly, with $\tauqfade$ being 0.8 to $0.2 \gyr$ from $\mstar = 5 \times 10 ^ 9$ to $2 \times 10 ^ {11} \msun$.
A lack of star formation implies a lack of dense, molecular gas, so it is interesting to compare $\tauqfade$ to observed molecular gas depletion times, defined as $M_{\rm H_2} / \sfr$, for galaxies near the quenching threshold.
\citet{SaiKauWan11} examined the molecular gas depletion times in a large sample of SDSS galaxies, finding typically $\sim 1 \gyr$ at $\mstar = 10 ^ {10 - 11} \msun$.
However, they found that the depletion time increases with decreasing SSFR, being $\sim 2 \gyr$ at $\ssfr \sim 10 ^ {-11} \yrinv$.
Such a long depletion time in galaxies near the quenching threshold is difficult to understand in a scenario in which satellites simply use up their molecular gas, which may suggest that additional processes are at play in reducing the molecular gas density in satellites as they are quenched, possibly via tidal or ram-pressure stripping or internal feedback processes.
However, note that the sample in \citeauthor{SaiKauWan11} is composed primarily of central (`field') galaxies, and it is unclear whether the significant gas reservoirs in these nearly quiescent galaxies result from them not fully depleting their cold gas while quenching, or by them accreting gas after they have quenched.
If the latter holds, the cold gas properties of satellites as they are being quenched could be quite different.

Several works have examined the gas content of satellites in the Virgo cluster, finding that while the atomic gas masses of Virgo satellites are significantly lower than for galaxies of the same mass in the field \citep{HuaHayGio12, SerOosMor12}, the molecular gas masses are quite similar \citep{KenYou86, YouBurDav11}.
This result suggest that, while processes like tidal or ram-pressure stripping may play a role in removing atomic gas from the outer regions of satellites, they have little impact on the molecular gas that fuels star formation.
Indeed, observations of ram-pressure stripping \citep{SunDonVoi07, ChuvGoKen09, AbrKenCrow11} typically show diffuse, atomic gas being stripped from the outer regions of the disc, while the dense, molecular gas towards the core survives intact, a phenomenon also seen in simulations \citep{TonBry09}.

Overall, we conclude that the gas reservoir in/around satellites at their time of infall is sufficient to fuel their necessary star formation histories and stellar mass growth.
Our result that satellite quenching can be parametrized simply by time since first infall, with no significant dependence on host halo mass, suggests that simple gas depletion (`strangulation') most naturally explains satellite quenching, though more work is needed to understand if gas self-depletion alone can account for our `delayed-then-rapid' quenching scenario with sufficiently short $\tauqfade$.
Our $\tauqfade$ values are marginally-to-significantly shorter than observed molecular gas depletion times, particularly for galaxies near the quenching threshold, possibly suggesting that some process other than simple molecular gas depletion is at play, though it is not clear if external stripping processes can explain this.
In \citetalias{WetTinCon13b}, we will examine these issues in more detail by developing physical models for satellite SFR evolution.

\subsection{Comparison with other work}

Our satellite quenching timescales are broadly consistent with previous works that parametrized the evolution of SFR in satellites and argued that it is affected over long ($2 - 3 \gyr$) timescales \citep[e.g.,][]{BalNavMor00, WanLiKau07, McGBalBow09, MahMamRay11}.
Recently, \citet{DeLWeiPog12} examined the infall times of satellites in a SAM applied to the Millennium simulation, accounting for hierarchical halo growth and group preprocessing; comparing with observed satellite quiescent fractions as a function of host halo mass and halo-centric radius, they argued that satellites take $\sim 5 - 7 \gyr$ to quench after falling into halos $> 10 ^ {13} \msun$.
However, these previous works generally only used observed quiescent/red fractions to constrain a single, overall quenching timescale, as in our Fig.~\ref{fig:quench-time_v_m-star}.
A significant aspect of our approach is using the overall SSFR distribution to constrain satellite SFR evolution more completely, through which we have shown that satellites experience a `delayed-then-rapid' quenching scenario, which is not possible measuring just quiescent fractions.
Furthermore, our use of empirically based satellite initial SFRs and a mock simulation group catalog to compare robustly with observations allows us to constrain these timescales empirically and robustly.

Our results place strong constraints on semi-analytic approaches to modeling the physics of satellite SFR evolution \citep[e.g.,][]{FonBowMcC08, KanvdB08, WeiKauvdL10, KimYiKho11}.
Our results suggest that a successful physical model would allow satellite SFR to evolve, environmentally unaffected, for $2 - 4 \gyr$ (depending on stellar mass) after infall, possibly through continued accretion/cooling of extended subhalo gas.
Our results also suggest that one does not need to impose any explicit dependence on host halo mass to this process.

We emphasize that our quenching timescales are valid for satellites at $z \sim 0$, given that our approach is sensitive to satellite SFR evolution within the last $\sim 4 \gyr$, and it is not clear that these timescales remain fixed at higher redshift.
We have checked this in our framework by applying our quenching timescales from $z = 0$ to satellites at higher redshift and comparing with the observed evolution in the quiescent fraction for all galaxies from Fig.~\ref{fig:qu.frac_v_z}.
While our model does agree within observational uncertainties at $z \lesssim 0.3$, we find that this approach quenches too few galaxies at higher $z$.
One one hand, this discrepancy could be interpreted as evidence against the accuracy of our 'delayed-then-rapid' quenching scenario.
Alternately, satellite quenching times simply may be shorter at higher redshift.
Using halo occupation modeling of galaxy spatial clustering measurements, \citet{TinWet10} showed that the satellite quiescent fraction does not evolve with redshift at fixed magnitude.
This lack of satellite evolution is supported broadly by observations of massive galaxies ($\mstar > 3 \times 10 ^ {10} \msun$) in X-ray-selected groups of mass $10 ^ {13 - 14} \msun$ out to $z \sim 1$ in COSMOS \citep{GeoLeaBun11}.
If the satellite quiescent fraction does not evolve at fixed satellite and host halo mass, then the ratio of a satellite's quenching time to its dynamical friction lifetime must remain roughly fixed, implying that the satellite quenching time is shorter at higher redshift: $\tq \propto (1 + z) ^ {-3 / 2}$ \citep{TinWet10}.
Furthermore, because the SSFR distribution of galaxies in groups is strongly bimodal out to at least $z \sim 0.4$ \citep{McGBalWil11}, this suggests that the delayed-then-rapid quenching scenario remains true at higher redshift.
Also, if $\tqdelay$ decreases more quickly than $\tauqfade$, then there would be a higher fraction of satellites at intermediate SSFRs at higher redshift, a trend that is suggested by the significant fraction of `green valley' galaxies in groups at $z \sim 1$ \citep{BalMcGWil11}.
In all, these result suggest that satellite quenching times are shorter at higher redshift, as we will investigate and quantify further in \citetalias{WetTinCon13b}.

We also emphasize that our results are predicated on the accuracy of the SHAM method for assigning stellar mass to both satellite and central subhalos across $z = 0 - 1$.
While numerous works support the (statistical) accuracy of this approach, as outlined in \S\ref{sec:galaxy_catalog_sim}, to any extent that SHAM assigns biased stellar masses to satellite subhalos at these redshifts, this would bias our derived quenching timescales.

We have argued that group preprocessing is the primary reason that satellites in more massive host halos are more likely to be quiescent.
Group preprocessing should manifest itself via an increased quiescent fraction for satellites that remain in a sub-group after falling into a cluster \citep{WhiCohSmi10}, to the extent that such sub-groups are observationally identifiable \citep{Coh12}.
Promisingly, examining the red fraction of galaxies in clusters at $z = 0.2 - 0.5$, \citet{LiYeeEll09} saw that, within and at the outskirts of these clusters, galaxies that appear associated with groups exhibit a higher red fraction than those that are not, providing direct evidence for the importance of group preprocessing.

Our results also connect with satellites of much lower mass in the Local Group.
Naively extending our quenching timescales in Fig.~\ref{fig:quench-time_v_m-star} to lower mass implies that dwarf satellites in the Milky Way take \tit{at least} $5 \gyr$ to quench after infall.
Given that the vast majority of satellites in the Local Group are quiescent, this implies that they fell in $> 5 \gyr$ ago, as supported by detailed comparisons of their positions and velocities to those of similar satellites in simulation \citep{RocPetBul12}.
Conversely, our quenching timescales also reinforce recent claims that the Large and Small Magellanic Clouds, which are both actively star-forming, fell into the Milky Way halo within the last few Gyrs \citep{BesKalHer07}: had they fallen in much earlier, our results indicate that they would no longer be actively star-forming.

Examining the stellar age distributions of satellites in the Local Group, \citet{OrbGneWei08} found that, even for those that are currently quiescent, a large fraction have experienced a significant amount of recent star formation: half of satellites with $\mstar > 10 ^ 7 \msun$ have formed more than 10\% of their stellar mass in the last $2 \gyr$.
This trend, observed in lower mass satellites than we examined, supports our general result that satellites continue to form stars over extended timescales and thus grow in stellar mass considerably after infall.
Examining more massive ($\mstar > 10 ^ 9 \msun$), quiescent elliptical/lenticular galaxies in the Coma cluster, \citet{TraFabDre08} found that their mean ages are identical to those of similar galaxies that are in the field, and that they have experienced star formation as recently at $z \sim 0.2$, trends that again suggest significant star formation after infall.
Finally, \citet{SmiLucPri12} examined the stellar ages of quiescent satellites in the Coma cluster, finding a significant decrease in age with cluster-centric radius at low mass ($\mstar < 10 ^ {10} \msun$) but a much weaker trend at higher mass.
As they argued, this trend implies that satellite-specific quenching plays a stronger role in quenching lower mass satellites, consistent with our results that more satellites were quenched as central galaxies prior to infall at higher mass.
Adding a large sample of stellar ages \citep[e.g.,][]{GalChaBri05} directly to our group catalog would provide additional constraints on our derived star formation histories, as we will examine in future work.

\section*{Acknowledgments}

We thank Michael Blanton, David Hogg, and collaborators for publicly releasing the NYU VAGC, Jarle Brinchmann and the MPA-JHU collaboration for publicly releasing their spectral reductions, and Martin White for simulation data.
We thank Marla Geha, Eyal Neistein, and Gary Mamon for enlightening discussions.
The simulation was analyzed at the National Energy Research Scientific Computing Center.

\bibliography{biblio}

\appendix

\section{Implications of satellite stellar mass growth for subhalo abundance matching} \label{sec:mass_growth_sham}

\begin{figure}
\centering
\includegraphics[width = 0.99 \columnwidth]{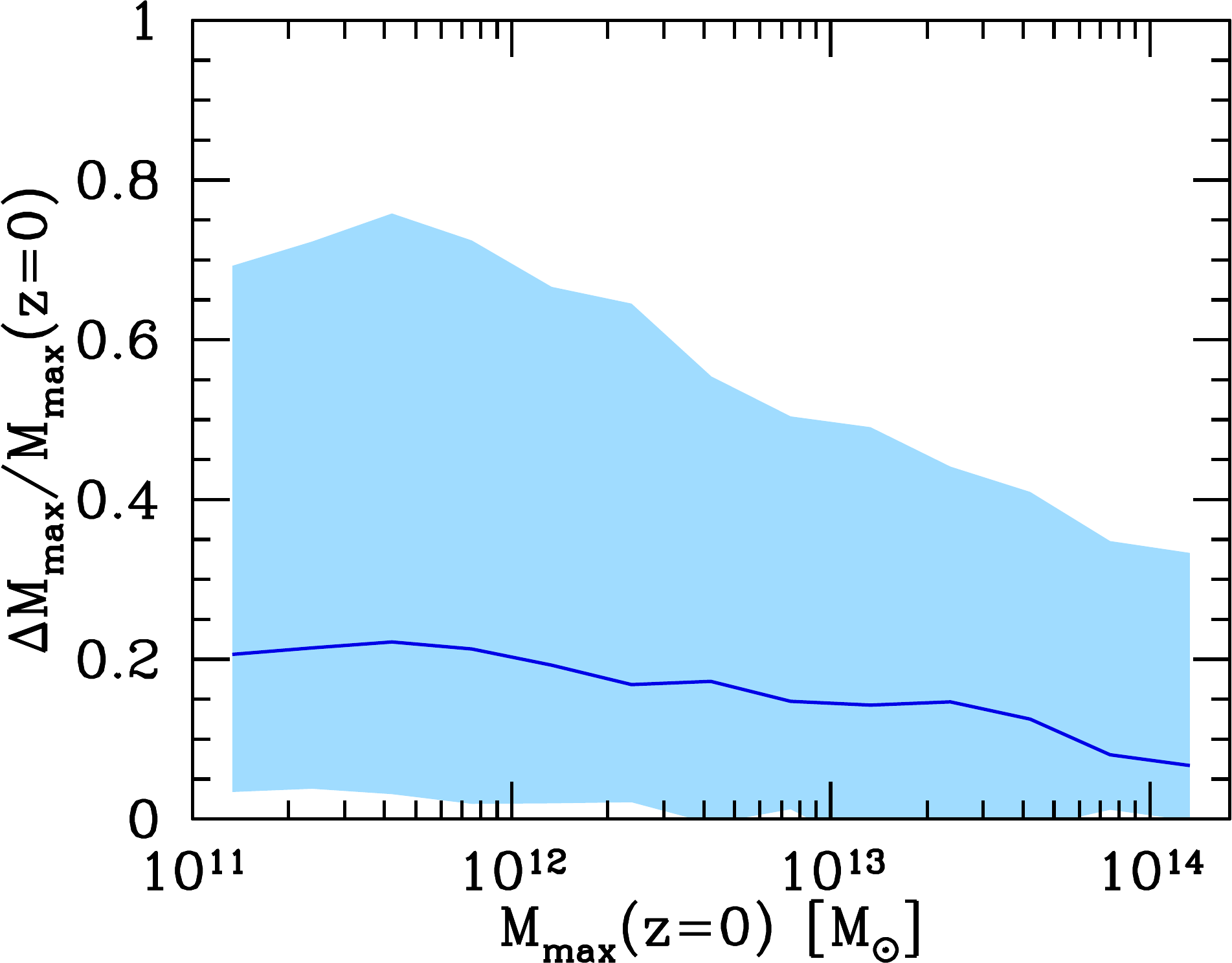}
\includegraphics[width = 0.99 \columnwidth]{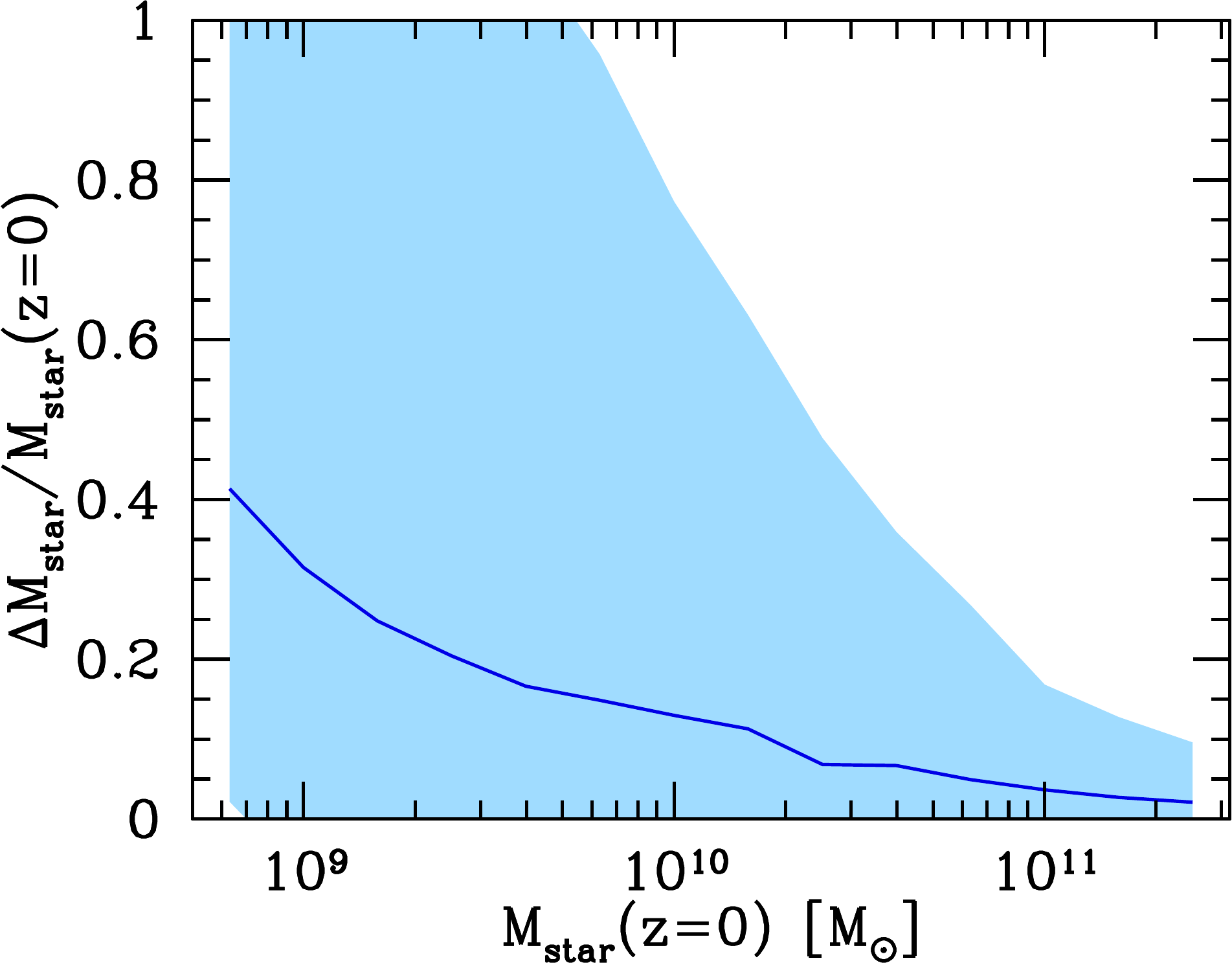}
\caption{
Fraction by which a central subhalo's $\mmax$ (top) and $\mstar$ (bottom) at $z = 0$ is higher than that of a satellite, for subhalos that had the same mass prior to the satellite's infall, according to SHAM.
Solid curve shows median value and shaded region shows 68\% distribution interval.
A satellite typically has 10 - 20\% lower $\mmax$ at $z = 0$ than it would have if it had remained a central, with considerable scatter.
At high mass, $\mstar$ growth is largely unaffected, but at low mass a satellite's $\mstar$ is systematically 40\% lower than for its central counterparts.
} \label{fig:m-grow_v_m_sham}
\end{figure}

In \S\ref{sec:m-star_growth}, we showed that stellar mass growth via star formation is nearly identical in satellite and central galaxies.
We now explore the implications of this result for subhalo abundance matching (SHAM), and we discuss other processes that influence the evolution of satellite stellar mass.

As described in \S\ref{sec:galaxy_catalog_sim}, the SHAM technique for assigning stellar mass to subhalos has been successful in matching many observed galaxy statistics.
At some level, this success is surprising: despite the differing physics of satellite and central galaxy evolution, with SHAM one assigns stellar mass to both a central and satellite subhalo based simply on the maximum subhalo mass (or circular velocity) that it experienced, $\mmax$, regardless of whether it is a central or satellite subhalo or its time since infall.
Our result, that satellite and central galaxies grow in stellar mass by essentially the same amount, on average, justifies assigning stellar mass to both satellite and central subhalos under a single, simple prescription.

However, our results suggest tension with the \tit{way} that SHAM typically is implemented, through the use of $\mmax$ (or maximum circular velocity).
Because satellite subhalos are stripped of their mass after falling into a host halo, their $\mmax$ almost always occurs prior to infall and remains constant thereafter.
(In our tracking scheme, a satellite can grow in $\mmax$ if it merges with another satellite.)
By contrast, a central subhalo's $\mmax$ continues to grow as its halo grows.
Thus, if one uses the same instantaneous $\mstar - \mmax$ relation to assign stellar mass to both satellite and central subhalos, which is how SHAM typically is applied, this necessarily implies that satellites have grown less in stellar mass than central subhalos.
Instead, the results of \S\ref{sec:m-star_growth} suggest that satellite subhalos should have a higher instantaneous $\mstar$ for their given $\mmax$ than central subhalos.

Under the assumption (for now) that satellite and central galaxies grow in stellar mass by the same amount, on average, we quantify this possible tension by examining the differential $\mmax$ growth of satellite versus central subhalos.
For each satellite at $z = 0$, we record the time that it reached $\mmax$ as a central subhalo, which typically was prior to first infall.
We then identify all central subhalos at $z = 0$ (discounting ejected satellites) that experienced the same $\mmax$ at the same time, and we compute the median difference in $\mmax$ at $z = 0$ between the satellite and these central subhalos, which we call $\Delta \mmax$.
We can then treat satellite and central subhalo stellar mass growths as self-consistent and identical by increasing each satellite subhalo's $\mmax$ at $z = 0$ by its $\Delta \mmax$.
Using these increased $\mmax$ values for satellites, we reassign stellar mass via SHAM to the whole subhalo population.
The difference between these stellar masses and the stellar masses from standard $\mmax$, which we call $\Delta \mstar$, indicates the amount by which the standard SHAM implementation underestimates satellites' stellar mass.

Fig.~\ref{fig:m-grow_v_m_sham} shows $\Delta \mmax / \mmax$ and $\Delta \mstar / \mstar$ as a function of satellites' mass at $z = 0$.
(For this exercise, we extend to masses below our SDSS group catalog limit).
The top panel shows the fraction by which a central subhalo's $\mmax$ at $z = 0$ is higher than that of a satellite, for subhalos that had the same $\mmax$ prior to the satellite's infall.
At low mass, a satellite's $\mmax$ at $z = 0$ is typically 20\% lower than its central subhalo counterparts, though with scatter to much higher values.
This fraction declines with increasing $\mmax$, but only weakly because of the competing effects that more massive central subhalos grow in mass more rapidly, but more massive satellites fell in more recently.

Fig.~\ref{fig:m-grow_v_m_sham} (bottom) shows the same, but for stellar mass, across a range that corresponds to the top panel.
The mass dependence is now stronger, because of the changing steepness in the $\mstar - \mmax$ relation with mass: at $\mmax \gtrsim 10 ^ {12} \msun$, $\mstar$ increases slowly with $\mmax$, so a given increase in $\mmax$ leads to a much smaller increase in $\mstar$, while at lower mass the opposite trend occurs.
The underestimation of satellite stellar mass is negligible at high mass, but at $\mstar < 10 ^ 9 \msun$ it is as high as 40\%, with scatter to much higher values.
Thus, to the extent that satellite and central galaxies mass growths are in fact identical, the standard implementation of SHAM significantly underestimates the stellar mass of low-mass satellites.

\citet{PasGalFon10} also examined the differential stellar mass growths of central and satellite galaxies since infall, though in the context of the semi-analytic model of \citet{WanDeLKit08}.
They also saw a negligible difference at high mass and an increasing difference at lower mass.
However, their difference was driven primarily by the rapid quenching of satellites after infall in their model, which we have argued cannot match the observed SSFR distribution \citep[see also][]{WeivdBYan06b, FonBowMcC08}.
Our empirically motivated approach indicates that the stellar mass growths via star formation should be nearly identical, on average, implying that satellite and central subhalos lie on separate instantaneous $\mstar - \mmax$ relations, a scenario that can, in principle, still match spatial clustering measurements \citep{NeiLiKho11}.
Promisingly, \citet{YanMovdB12} recently implemented a self-consistent approach to SHAM, in which they found evidence that satellite and central subhalo stellar mass growths are similar, in support of our results.

However, an important caveat to Fig.~\ref{fig:m-grow_v_m_sham} is that star formation is not the only process that affects satellite stellar mass evolution.
Satellites can merge with one another \citep{AngLacBau09, KimBauCol09, WetCohWhi09a, WetCohWhi09b}, though we expect that the impact of satellite mergers on stellar mass growth is subdominant at these masses (\citetalias{Wet12}).
More importantly, satellites can lose stellar mass via tidal stripping.
As outlined in \S\ref{sec:simulation}, we use a simple binary procedure for merging/disrupting highly stripped satellite subhalos, but we do not account for partial stripping of stellar mass.
There are many reasons to expect that surviving satellites have been at least partially stripped, based on both the properties of satellites \citep[e.g.,][]{YanMovdB09, KimBauCol09, PasGalFon10, SimWeiDav10} and the presence of diffuse intracluster light (ICL) from stripped satellites \citep[e.g.,][]{WilGovWad04, ConWecKra07, RudMihFre09, PucSprSij10}.
It is possible that the canceling effects of stellar mass growth via star formation and stellar mass loss via tidal stripping cause satellite and central subhalos to lie on essentially the same instantaneous $\mstar - \mmax$ relation \citep[see also][]{SimWeiDav10}.
If true, then SHAM ascribes the correct stellar masses to satellites though a fortuitous coincidence: the retarded growth of a satellite's $\mmax$, and thus its $\mstar$, accurately captures the physical process of tidal stripping.
If so, then Fig.~\ref{fig:m-grow_v_m_sham} must also indicate the average amount of stellar mass stripping that surviving satellites at $z = 0$ have suffered.
Direct observations of the stellar mass fraction in ICL are $\sim 10 - 25\%$ \citep{ZibWhiSch05, GonZarZab07, KriBer07}, broadly consistent with Fig.~\ref{fig:m-grow_v_m_sham}.

Finally, we note the implications of these results to SHAM as applied to dwarf satellites in the Local Group.
While Fig.~\ref{fig:m-grow_v_m_sham} extends only down to about the mass of the Small Magellanic Cloud, it also indicates the increasing importance of satellite stellar mass growth at lower mass.
Thus, understanding the relative importance of stellar mass growth and tidal stripping becomes even more important for dwarf satellites in the Local Group.
If local dwarf satellites have not been stripped of stellar mass significantly, as might be implied by their tight luminosity-metallicity relation \citep{KirCohSmi11}, then Fig.~\ref{fig:m-grow_v_m_sham} implies that the usual SHAM approach may underestimate their stellar masses.

\label{lastpage}

\end{document}